\documentclass[journal]{IEEEtran}

\usepackage{cite,csquotes}
\usepackage[breaklinks=true]{hyperref}
\usepackage[cmex10]{amsmath}

\usepackage{verbatim}
\usepackage{amssymb}
\usepackage{amsthm}
\newtheorem{theorem}{Theorem}
\newtheorem{lemma}{Lemma}

\newtheorem{example}{Example}
\usepackage{mathtools}
\usepackage{commath}
\usepackage{dsfont}

\newtheoremstyle{break}
  {\topsep}{\topsep}%
  {\itshape}{}%
  {\bfseries}{}%
  {\newline}{}%
\theoremstyle{break}

\theoremstyle{definition}

\usepackage{tikz,adjustbox}
\usetikzlibrary{shapes}

\title{An Efficient and Incentive-Compatible Mechanism for Energy Storage Markets}
\author{Bharadwaj Satchidanandan and Munther A. Dahleh}

\begin{document}

\maketitle

\begin{abstract}
A key obstacle to increasing renewable energy penetration in the power grid is the lack of utility-scale storage capacity. Transportation electrification has the potential to overcome this obstacle since Electric Vehicles (EVs) that are not in transit can provide battery storage as a service to the grid. This is referred to as EV-Power grid integration, and could potentially be a key milestone in the pathway to decarbonize the electricity and the transportation sectors. We first show that if EV-Power grid integration is not done carefully, then contrary to improving the cost efficiency of operating the grid, it could in fact be counterproductive to it. This fundamentally occurs due to two phenomena operating in tandem -- the randomness of EV usage patterns and the possibility of strategic behavior by EV operators.
We present a market-based solution to address this issue. Specifically, we develop a mechanism for energy storage markets using which the system operator can efficiently integrate a fleet of strategic EVs with random usage patterns into the grid, utilize them for storage, and satisfy the demand at minimum possible cost.
\end{abstract}


\section{Introduction}\label{introduction}
A major impediment to high renewable energy penetration in the power grid is the scarcity of energy storage capacity in the grid. Utility-scale battery storage is expensive at current technology, and so any energy that is generated must be consumed immediately. This paradigm could change substantially with increased Electric Vehicle (EV) penetration since EVs that are not in transit can provide battery storage as a service to the grid. Prior studies estimate that on an average, a car is parked for more than $95\%$ of the time \cite{carsParked}, indicating the huge potential for EVs to double as energy storage resources in the grid. As an illustration, take the example of the state of Massachusetts. It consumes an average of $146$GWh of electric energy per day \cite{MAelectricity}. On the other hand, the battery capacity of a Tesla Model S EV is about $100$kWh. This implies that about $1.4$ million EVs possess enough battery capacity to power Massachusetts for an entire day. This amounts to less than $64\%$ of the vehicles registered in Massachusetts today \cite{MAvehicles}. The situation is similar in most other parts of the US and the world, indicating that even moderate levels of EV penetration could provide significant storage capacity.

The time periods during which an EV can lease its battery to the grid are private knowledge of the EV operator. In particular, it is unknown to the Independent System Operator (ISO). However, the ISO requires this information to optimally operate the grid, or more precisely, to determine the optimal power dispatch of the generators and the optimal storage schedule of the EVs. Consequently, the ISO requests the EV operators to report to it in the day-ahead market the time periods during which they can lease their battery the following day. However, this brings forth two challenges that need to be addressed. 

The first challenge is that the travel times of people are in general random, and so the EV operators may not precisely know in the day-ahead market the time periods during which they can lease their batteries the following day. Rather, they may know these time periods only with some uncertainty. To account for this, we model the time periods during which an EV can lease its battery as a random variable, and require that the EV operators only report the probability distribution of this random variable in the day-ahead market. 

The second challenge is that the EV operators could be strategic, and so they may not report the aforementioned probability distribution truthfully. As we elaborate in Section \ref{problemFormulation}, each EV operator has associated with it a utility function, and the EV operators bid strategically so as to maximize their respective utilities. Moreover, having bid some probability distribution in the day-ahead market, an EV may not remain connected to the grid until its deadline the following day if there is possibility for it to obtain a higher utility by doing so than by disconnecting at its deadline. Such behavior could potentially be counterproductive to the cost- and energy-efficient operation of the grid. The following example illustrates this issue.

\begin{example}
Suppose that a day consists of two time periods, and suppose that the demand sequence $\mathbf{d}$ of the load in these time periods is $\mathbf{d}=\{0,1\}.$ Let the production function $c^g$ of the generator be such that $c^g(\{1,0\})=0$ and $c^g(\{0,1\})=2.$ That is, it costs the generator $0$ to produce $1 \mathrm{J}$ of energy at time period $1$ and $0\mathrm{J}$ of energy at time period $2$, and so on. The cost $c^g$ for all other $2$-tuples is infinite. We suppose that this generator has a low ramping rate -- a characteristic that is typical of high-efficiency generators -- and so its power dispatch must be scheduled well in advance of the time of power delivery. Specifically, its power dispatch must be scheduled in the day-ahead market. 

The system also consists of a reserve generator which has a high ramping rate which can produce and sell energy in the spot market to balance real-time demand-supply mismatches. Let the production function $c^s$ of the reserves be such that $c^s(\{0,0\})=0$ and $c^s(\{0,1\})=11.$ The cost $c^s$ for all other $2$-tuples is infinite.

Suppose that there is only one EV in the system with a battery capacity of $1\mathrm{J}$. As we elaborate in Section \ref{problemFormulation}, the usage pattern of an EV on any given day is specified by a quantity known as its ``deadline" on that day. An EV's deadline on a given day is defined as the time period until which the EV can lease its battery to the grid on that day. Then, the usage pattern of an EV being random is equivalent to its deadline being random. Suppose that the EV's deadline takes the value $1$ with probability $p$ and the value $2$ with probability $1-p$. 

As elaborated in Section \ref{problemFormulation}, each EV has a cost function associated with it. Suppose that the cost incurred by an EV is equal to the negative of the net energy injected into it during the time that it is connected to the grid.

Now, the ISO is confronted with two options to meet the demand. The first option is for it to schedule the generator to produce the energy sequence $\mathbf{g}=\{0,1\}$ and use it to serve the load. This results in a total cost of meeting the demand --- defined as the sum of the costs incurred by the generator, the EV, and the reserves --- to be equal to $2$. 

The second option is to schedule the generator to produce the energy sequence $\mathbf{g}=\{1,0\}$ and store the energy generated in the first time step in the EV. If the EV remains connected to the grid in the second time step, then the ISO discharges it to satisfy the demand, resulting in a total cost of $0$. On the other hand, if the EV disconnects at time step $1$, then the ISO purchases $1\mathrm{J}$ in the spot market at time step $2$ at cost $c^s(\{0,1\})=11$ to satisfy the demand, thereby resulting in a total cost of $10$. Hence, if the EV disconnects at its true deadline, then the total cost of meeting the demand equals $0$ with probability $1-p$ and equals $10$ with probability $p$. Hence, the total expected cost of meeting the demand if the ISO decides on the second option is equal to $10p$.

Now, the goal of the ISO, as we elaborate in Section \ref{problemFormulation}, is to minimize the total expected cost of meeting the demand, and so it must choose the first option if $2\leq 10p$ and the second option if $2>10p.$ However, the difficulty is that the ISO does not know the value of $p,$ and must rely only on the value $\widehat{p}$ reported by the EV in the day-ahead market in order to make the decision. The EV bidding $\widehat{p}=p$ is not a dominant strategy. To see this, suppose that $p=0.21.$ If the EV bids $\widehat{p}=0.21,$ then the ISO would decide on the first option, and so the cost incurred by the EV would be equal to $0.$ On the other hand, if the EV bids $\widehat{p}=0.19,$ then that causes the ISO to decide on the second option, thereby resulting in the EV being charged with $1\mathrm{J}$ in the first time step. If the EV then disconnects at its deadline, then it would exit the system with a charge of $1\mathrm{J}$ with probability $0.21$ and a charge of $0\mathrm{J}$ with probability $0.79.$ This would result in it incurring an average cost of $-0.21$, which is lesser than the average cost of $0$ that it would incur if it bids $p$ truthfully. However, the total expected cost of meeting the demand as a result of the EV's false bid is equal to $2.1,$ which is greater than the total cost of $2$ that would result if the ISO simply decides to never utilize the EV for storage.

The above example illustrates a scenario wherein the EV can lower its average cost by misreporting only its deadline distribution. However, the EV could also misreport its deadline realization to lower its cost. To see this, consider the case when $p=0.19$ and suppose that the EV reports $\widehat{p}=p$ truthfully in the day-ahead market. It follows from the above discussion that the ISO would decide on the second option. Now, if the EV reports its deadline truthfully in real time, then the expected cost that it would incur is equal to $-0.19$. On the other hand, by misreporting its deadline realization to be equal to time step $1$, the EV can exit the system with $1\mathrm{J}$ of charge, thereby resulting in it incurring a lower cost of $-1$. Since the ISO is not privy to the EV's deadline realization, it cannot ascertain if the EV disconnects at the first time step due to its deadline arriving at that time or due to strategic behavior. 
The total expected cost of meeting the demand as a result of the EV's false report would equal $10,$ which is greater than both the total expected cost of $1.9$ that would result if the EV reports its deadline realization truthfully, and also the total cost of $2$ that would result if the ISO simply decides to never utilize the EV for storage. 
\end{example}

The above example illustrates how strategic EV behavior
not only defeats the purpose of utilizing EVs as energy storage units, but could also be counterproductive to the cost- and energy-efficient operation of the grid since it could potentially result in real-time supply shortages which in turn increases the ISO's dependence on the expensive and energy-inefficient reserves. Therefore, if EVs are to be efficiently integrated for storage, it is imperative to devise incentive structures that drive EV operators towards truthful behaviors. Specifically, it is necessary to devise mechanisms that (i) incentivize the EVs report the probability distribution of their usage patterns truthfully in the day-ahead market, so that the ISO can optimally plan the power dispatch and storage schedules, and (ii) incentivize the EVs to remain plugged into the grid until their actual deadlines, so that there are no untoward supply shortages in real time. In this paper, we present a mechanism that achieves both of these objectives.

At a high level, the mechanism consists of a decision rule that specifies an optimal power dispatch sequence of the generator and an optimal energy storage policy for each EV as a function of the deadline distributions that the EVs report in the day-ahead market, and a payment rule that incentivizes EVs to report their deadline distributions \emph{and} deadline realizations truthfully. The payment rule consists of two components --- (i) a ``day-ahead payment" that reflects the expected cost savings in operating the grid due to the storage opportunity that the EVs are expected to provide as per their reported deadline distributions, and (ii) a carefully designed ``end-of-the-day settlement" that adjusts the transfers meted out to each EV based on a comparison of what it had reported in the day-ahead market and the actual departure profiles of the EVs in real time. One of the functionalities of the end-of-the-day settlement is to penalize EVs for deviations of their empirically observed behavioral patterns from what is expected as per the probability distributions that they report in the day-ahead market. We show how the composite payment rule renders truthful bidding in both the day-ahead market and in real time a dominant strategy for every EV, thereby enabling the ISO to satisfy the demand at minimum possible cost. 
To the best of our knowledge, we are unaware of any other work that addresses the problem of integrating a fleet of \emph{strategic} EVs with \emph{random usage patterns} into the grid, encompassing both engineering and economic aspects of the composite system. 


\noindent\textbf{Notation:} Given a vector $\mathbf{x},$ we denote by $x(i)$ the $i^{th}$ component of $\mathbf{x}.$ Given a vector $\mathbf{x}(\boldsymbol{\theta})$ which is a function of the variable $\boldsymbol{\theta}$, we denote its $i^{th}$ component by $x(i;\boldsymbol{\theta})$. Given a vector $\mathbf{x},$ we denote by $\mathbf{x}_{-i}$ the vector $\mathbf{x}$ with its $i^{th}$ component removed, and by $[y,\mathbf{x}_{-i}]$ the vector whose $i^{th}$ component is $y$ and the other components are $\mathbf{x}_{-i}.$  Given a sequence $\{\mathbf{x}(1),\mathbf{x}(2),\hdots\},$ we use $\mathbf{x}^l$ to denote the $l-$length sequence $\{\mathbf{x}(1),\hdots,\mathbf{x}(l)\},$ and $\mathbf{x}^\infty$ to denote the entire sequence. We occasionally use $[x_i^\infty,\mathbf{x}_{-i}^\infty]$ to denote $\mathbf{x}^\infty$, especially when it is necessary to draw attention to the $i^{th}$ component of the vectors. We denote by $\mathds{1}_{\{\mathrm{Statement}\}}$ the indicator function that takes the value $1$ if $\mathrm{Statement}$ is true and takes the value $0$ otherwise. 

\section{Related Work}\label{relatedWork}

Technologies that utilize EVs as energy storage resources are broadly referred to as vehicle-to-grid technologies, and a large body of literature exists on this topic. Reference \cite{v2gFeasibilityStudy} provides a feasibility study of vehicle-to-grid systems. Reference \cite{Gross1} presents a framework for vehicle-to-grid implementation. Among other aspects, it recognizes the need for incentive mechanisms to ensure adequate participation of EV operators. Mechanisms to elicit private valuations of EVs under a variety of settings and to achieve various objectives are presented in \cite{Jennings1,Nejad2017,Hiskens,PSP1,SaadBasar,ACC2021} among many other papers. However, most of the existing literature fail to model at least one, if not all, of the following aspects of EVs: (i) the heterogeneity of EV deadlines, (ii) the stochasticity of EV deadlines, (iii) the storage opportunity provided by EVs, and (iv) the possibility of strategic behavior by EV operators. The concoction of aspects (ii) and (iv) demands particular attention since the stochastic nature of EV deadlines provides EVs significant leeway for strategic behavior since the grid operator cannot ascertain whether an EV's departure at any given time on any given day is a consequence of its deadline arrival or is a consequence of strategic behavior. In this paper, we model all of the aforementioned aspects in a holistic framework and address them thoroughly.

\section{Problem Formulation}\label{problemFormulation}

Consider a power system with a single generator, a single load, and $n_s$ storage units or EVs. In addition to EVs, the storage units could also include devices such as Powerwalls \cite{powerwall} that individual households and firms could have installed. 
We divide time into days, and divide a day into $T$ time intervals. Denote by $d_l(t)$ the energy demand of the load on the $l^{th}$ day at time $t,$ $l\in\mathbb{Z}_+$ and $t\in\{1,\hdots,T\}.$ The demand sequence $\mathbf{d}_l\coloneqq \big[d_l(1),\hdots,d_l(T)\big]$ is a random variable and can typically be forecast in the day-ahead market to an accuracy of within $5\%$ \cite{DAestimationError}. However, in order to minimize clutter and expose the main ideas clearly, we assume that it is known exactly in the day-ahead market, and furthermore, that it remains the same on all days. Consequently, we drop the subscript $l$ and denote the demand sequence simply as $\mathbf{d}=[d(1),\hdots,d(T)].$ We describe in Section \ref{extensions} how this assumption can be relaxed. 

\subsection{EV Deadlines}\label{EVDeadlinesSubsection}
On any day $l,$ $l\in\mathbb{Z}_+,$ the time intervals in which an EV $i$ can lease its battery to the grid are characterized by a parameter $\delta_i(l)\in\{1,\hdots,T\}$ called EV $i$'s ``deadline" on day $l.$ An EV $i$ on day $l$ is said to have deadline $\delta_i(l)$ if it can lease its battery to the grid at all times lesser than or equal to $\delta_i(l)$ and incurs a large cost for remaining connected beyond time $\delta_i(l)$. 

\subsubsection{Deadline distributions}
The deadline $\delta_i(l)$ is modeled as a random variable, and we suppose that the deadline sequence $\{\delta_i(1),\delta_i(2),\hdots,\}$ is Independent and Identically Distributed (IID). Since $\delta_i(1)$ takes values from the set $\{1,\hdots,T\}$, the set of distributions that it can assume is parameterized by the $\mathbb{R}^{T}$- dimensional probability simplex $\overline{\Theta}$. We denote by $\theta_i\in\overline{\Theta}$ the parameter vector corresponding to EV $i$'s deadline distribution, and by $\mathbb{P}_{\theta_i}$ the probability distribution of the deadline. That is, for any $t\in\{1,\hdots,T\},$ the quantity $\mathbb{P}_{\theta_i}(t)$ denotes the probability that $\delta_i(1)$ equals $t.$ 

While in general, $\theta_i$ could take any value in $\overline{\Theta},$ we assume for certain technical reasons that there exists $\epsilon_{\theta}>0$ such that for all $i\in\{1,\hdots,n_s\}$ and for all $t\in\{1,\hdots,T\},$ $\mathbb{P}_{\theta_i}(t)\geq\epsilon_{\theta}.$ Consequently, given $\epsilon_{\theta},$ we define the set
\begin{align*}
    \Theta\coloneqq\{\theta\in\overline{\Theta}:\mathbb{P}_{\theta}(t)\geq\epsilon_{\theta}\;\; \mathrm{for\;all}\;\; t\in\{1,\hdots,T\}\},
\end{align*}
so that for all $i\in\{1,\hdots,n_s\},$
\begin{align}
    \theta_i\in\Theta.\label{thetainTheta}
\end{align}

We assume that the deadlines of different EVs are independent random variables so that the joint distribution of the EVs' deadlines on any given day is the product distribution $\mathbb{P}_{\theta_1}\times\hdots\times\mathbb{P}_{\theta_{n_s}}.$ We define $\boldsymbol{\theta}\coloneqq[\theta_1,\hdots,\theta_{n_s}].$

\subsubsection{Deadline realizations}
We suppose that for every $i\in\{1,\hdots,n_s\}$ and every $l\in\mathbb{Z}_+,$ the realization of $\delta_i(l)$ is drawn ``by nature" at the beginning of day $l$ according to $\mathbb{P}_{\theta_i}$ and is revealed to EV $i$ at the beginning of day $l.$ In particular, the realization of $\delta_i(l)$ is unknown in the day-ahead market corresponding to day $l.$ The details of the day-ahead market are described in later subsections.

\subsection{Cost functions}

We denote by $c^g:\mathbb{R}_{\geq0}^T\to\mathbb{R}$ the production function of the generator so that $c^g(\mathbf{g})$ is the cost incurred by the generator for producing the energy sequence $\mathbf{g}=\big[g(1),\hdots,g(T)\big]$ on any given day, where $g(t)$ denotes the amount of energy produced at time $t$.

Demand-supply mismatches that occur in real-time are typically compensated by the ISO by purchasing additional energy in the spot market. We denote by  $c^{s}:\mathbb{R}^T\to\mathbb{R}$ the production function of the reserves so that $c^{s}(\mathbf{g}_s)$ is the cost incurred by it if it produces the energy sequence $\mathbf{g}_s=[g_s(1),\hdots,g_s(T)].$. We allow for the reserves to also consume excess energy, and so a negative value of $g_s(t)$ denotes an absorption of $g_s(t)$ units of energy at time $t$, $t\in\{1,\hdots,T\}$. In case it is infeasible for the reserves to absorb energy in real time, the cost of $T-$tuples that contain negative entries are set to infinity. 

For any EV $i$, we mean by the term ``storage sequence of
the EV $i$ on day $l$" a $T-$length sequence that specifies the energy stored in EV $i$ at each time of the day. Note that the storage sequence of an EV uniquely specifies how much energy must be injected or consumed from the EV at each time of the day --- a decision that, as we will see shortly, the ISO must make for every EV for optimal operation of the grid. Every EV $i$ has associated with it a cost function $c_i^{EV}:\{1,\hdots,T\}\times\{1,\hdots,T\}\times\mathbb{R}^T\to\mathbb{R}$ that specifies the cost incurred by the EV on any day $l$ as a function of (i) its deadline $\delta_i(l)$ on that day, (ii) the actual time $\widehat{\delta}_i(l)$ at which it disconnects from the grid on that day, and (iii) its storage sequence $\mathbf{h}_{i,l}$ on that day, and is defined as
\begin{align}
    c_i^{EV}(\delta_i(l),\widehat{\delta}_i(l),\mathbf{h}_{i,l})=-h_{i,l}(\widehat{\delta}_i(l))&\mathds{1}_{\{\widehat{\delta}_i(l)\leq\delta_i(l)\}}\nonumber\\
    +&J_m\mathds{1}_{\{\widehat{\delta}_i(l)>\delta_i(l)\}},\label{EVCost}
\end{align}
where $J_m\in\mathbb{R}_{>0}$ denotes the cost incurred by the EV if it misses its deadline.
In other words, if EV $i$ has deadline $\delta_i(l)$ but disconnects from the grid only after its deadline, then it incurs a cost $J_m$ for missing its deadline, but on the other hand, if it disconnects from the grid before its deadline, then the cost that it incurs is proportional to the net energy injected into it, viz., $-h_{i,l}(\widehat{\delta}_i(l))$. {The above cost function models a situation in which the deadlines of EVs are hard constraints as defined in Section \ref{EVDeadlinesSubsection}. Relaxing this assumption would entail defining the cost for missed deadlines to gradually increase as a function of the duration by which the deadline is missed. }

\subsection{Storage Policy}

The stochasticity of EV departure times necessitates the ISO to devise a storage policy in order to determine the storage sequence of each EV on each day. The storage policy, in general, could be a randomized policy.
Specifically, a storage policy $\boldsymbol{\pi}$ is specified by a probability space $(\Omega_{\boldsymbol{\pi}},\mathcal{F}_{{\boldsymbol{\pi}}},\mathbb{P}_{\boldsymbol{\pi}})$ and collection of functions $\{\pi_1,\hdots,\pi_{n_s}\}$ where the $i^{th}$ function $\pi_i:\{1,\hdots,T\}\times\{1,\hdots,T\}^{n_s}\times\Omega_{\boldsymbol{\pi}}\to\mathcal[0,B_i]$ specifies the energy that must be stored in EV $i$ at each time of the day as a function of (i) the departure profiles of the EVs on that day, and (ii) a variable that serves as the source of randomness of the policy. More precisely, if $\widehat{\boldsymbol{\delta}}(l)\in\{1,\hdots,T\}^{n_s}$ denotes the vector of departure times of EVs on day $l$, then $\pi_i(t,\widehat{\boldsymbol{\delta}}(l),\omega_{\boldsymbol{\pi}}(l))$ specifies the energy that must be stored in EV $i$ at time $t$ on day $l$, where $\omega_{\boldsymbol{\pi}}(l)\in\Omega_{\boldsymbol{\pi}}$ is a random variable that the policy draws on day $l$ according to  $\mathbb{P}_{\boldsymbol{\pi}}$ independently of all other random variables realized until that day. 
We denote by
\begin{align*}
    \boldsymbol{\pi}_i(\widehat{\boldsymbol{\delta}}(l),\omega_{\boldsymbol{\pi}}(l))\coloneqq[\pi_i(1,\widehat{\boldsymbol{\delta}}(l),\omega_{\boldsymbol{\pi}}(l))\hdots\pi_i(T,\widehat{\boldsymbol{\delta}}(l),\omega_{\boldsymbol{\pi}}(l))]
\end{align*}
the storage sequence of EV $i$ on day $l$ as a function of the EVs' departure profiles $\boldsymbol{\widehat{\delta}}(l)$ and the ``coin flip" $\omega_{\boldsymbol{\pi}}(l)$ of the storage policy. 

For a storage policy $\boldsymbol{\pi}$ to be implementable by the ISO, it must not charge or discharge any EV after it disconnects from the grid. That is, the policy $\boldsymbol{\pi}$ must satisfy the condition that for every $i\in\{1,\hdots,n_s\}$, every $\widehat{\boldsymbol{\delta}}(l)\in\{1,\hdots,T\}^{n_s},$ and every $\omega_{\boldsymbol{\pi}}(l)\in\Omega_{\boldsymbol{\pi}},$ $$\pi_i(t,\widehat{\boldsymbol{\delta}}(l),\omega_{\boldsymbol{\pi}}(l))=\pi_i(\widehat{\delta}_i(l),\widehat{\boldsymbol{\delta}}(l),\omega_{\boldsymbol{\pi}}(l))$$ for all $t\geq\widehat{\delta}_i(l).$ We denote by $\Pi$ the set of all implementable storage policies.

\subsection{The Independent System Operator's Objective}

{Prior to each day, the ISO runs a day-ahead market in which it must decide the energy dispatch sequence of the generator and the storage policy of the EVs for that day. As mentioned before, the deadlines of the EVs for any given day realize only after the day-ahead market for that day closes, and so the only information on which the ISO can base its day-ahead market decisions are the deadline distributions that the EVs bid in the market. Suppose for a moment that the EVs are not strategic and that they bid the deadline distributions $\boldsymbol{\theta}$ truthfully. How should the ISO compute the energy dispatch sequence and the storage policy? Since the ISO does not know the deadline realizations in the day-ahead market, it chooses these quantities so as to minimize the expected cost of meeting the demand on the following day. This is elaborated in more detail below. Note that since the deadline distributions of the EVs are assumed to remain the same on all days, it suffices for the ISO to compute these quantities just once, namely, in the day-ahead market prior to day $1,$ and reuse them on all days. We describe in Section \ref{extensions} how to address the more general scenario wherein the deadline distributions could be different on different days.}

{Suppose that the ISO decides the generator's energy dispatch sequence to be $\mathbf{g}=[g(1),\hdots,g(T)]$ and the storage policy to be $\boldsymbol{\pi}$ in the day-ahead market. Then,
\begin{align}
    {{g}_s}(t,{\boldsymbol{\delta}}(l),\mathbf{g},\boldsymbol{\pi},\omega_{\boldsymbol{\pi}}(l))\;\;\;\;\;\;\;\;\;\;\;\;\;\;&\nonumber\\
    =d(t)-\bigg[g(t)+\sum_{i=1}^{n_s}\bigg(\pi_i(&t-1,{\boldsymbol{\delta}}(l),\omega_{\boldsymbol{\pi}}(l))\nonumber\\
    &-\pi_i(t,{\boldsymbol{\delta}}(l),\omega_{\boldsymbol{\pi}}(l))\bigg)\bigg]\label{gstDefn}
\end{align}
is the real-time demand-supply mismatch at time $t$ on day $l$. Hence, the ISO has to purchase the energy sequence 
\begin{align*}
    \mathbf{g}_s({\boldsymbol{\delta}}&(l),\mathbf{g},\boldsymbol{\pi},\omega_{\boldsymbol{\pi}}(l))\nonumber\\
    =&[{g}_s(1,{\boldsymbol{\delta}}(l),\mathbf{g},\boldsymbol{\pi},\omega_{\boldsymbol{\pi}}(l)),\hdots,{g}_s(T,{\boldsymbol{\delta}}(l),\mathbf{g},\boldsymbol{\pi},\omega_{\boldsymbol{\pi}}(l))]
\end{align*}
in the spot market on day $l$ at price $c^{s}({\mathbf{g}_s}({\boldsymbol{\delta}}(l),\mathbf{g},\boldsymbol{\pi},\omega_{\boldsymbol{\pi}}(l))).$ Therefore, the total cost of satisfying the demand on day $l$ -- defined as the sum of the costs incurred by the generator, the reserves, and the EVs -- is $$c^g(\mathbf{g})+c^{s}({\mathbf{g}_s}({\boldsymbol{\delta}}(l),\mathbf{g},\boldsymbol{\pi},\omega_{\boldsymbol{\pi}}(l)))-\sum_{j=1}^{n_s}\pi_j(\delta_j(l),\boldsymbol{\delta}(l),\omega_{\boldsymbol{\pi}}(l)).$$ Since $\boldsymbol{\delta}(l)$ and $\omega_{\boldsymbol{\pi}}(l)$ are random variables, this cost is a random variable for any energy dispatch sequence $\mathbf{g}$ and any storage policy $\boldsymbol{\pi}$ that the ISO chooses in the day-ahead market. The ISO's objective is to minimize the total expected cost of meeting the demand, and therefore it chooses the energy dispatch sequence $\mathbf{g}^*$ and the storage policy $\boldsymbol{\pi}^*$ as a solution to the stochastic program
\begin{align}
\underset{\mathbf{g}\in\mathbb{R}^T,\boldsymbol{\pi}\in\Pi}{\mathrm{Minimize}}\;\;\mathbb{E}_{(\boldsymbol{\delta},\omega_{\boldsymbol{\pi}})\sim\mathbb{P}_{\boldsymbol{\theta}}\times\mathbb{P}_{\boldsymbol{\pi}}}\Big[ c^g(\mathbf{g})+c^{s}&\big({\mathbf{g}_s}({\boldsymbol{\delta}},\mathbf{g},\boldsymbol{\pi},\omega_{\boldsymbol{\pi}})\big)\nonumber\\
-&\sum_{j=1}^{n_s}{{\pi}_j(\delta_j,\boldsymbol{\delta},\omega_{\boldsymbol{\pi}})}\Big].\label{ISO_asymptoticProblemLong}
\end{align}
We introduce a few definitions to compactify the notation. Define
\begin{align}
    \widehat{c^s}({\boldsymbol{\delta}},\mathbf{g},\boldsymbol{\pi})\coloneqq\mathbb{E}_{\omega_{\boldsymbol{\pi}}\sim\mathbb{P}_{\boldsymbol{\pi}}}\big[c^s(\mathbf{g}_s({\boldsymbol{\delta}},\mathbf{g},\boldsymbol{\pi},\omega_{\boldsymbol{\pi}}))\big],\label{cHatDefn}
\end{align}
which has the interpretation as the average cost of energy purchase in the spot market if the generator produces the energy sequence $\mathbf{g},$ the EVs disconnect at times $\boldsymbol{\delta},$ and the ISO employs the storage policy $\boldsymbol{\pi}.$ Here, the averaging is carried out over the random choices made by the storage policy $\boldsymbol{\pi}$. Similarly define
\begin{align}
    \widehat{\pi}_i(t,{\boldsymbol{\delta}})\coloneqq\mathbb{E}_{\omega_{\boldsymbol{\pi}}\sim\mathbb{P}_{\boldsymbol{\pi}}}[\pi_i(t,{\boldsymbol{\delta}},\omega_{\boldsymbol{\pi}})]\label{piHatDefn}
\end{align}
and let $\widehat{\boldsymbol{\pi}}_i({\boldsymbol{\delta}})\coloneqq\big[\widehat{\pi}_i(1,{\boldsymbol{\delta}}),\hdots,\widehat{\pi}_i(T,{\boldsymbol{\delta}})\big].$
Finally define 
\begin{align}
    \beta(\boldsymbol{\delta},\mathbf{g},\boldsymbol{\pi})\coloneqq c^g(\mathbf{g})+\widehat{c^s}(\boldsymbol{\delta},\mathbf{g},\boldsymbol{\pi})-\sum_{j=1}^{n_s}{\widehat{\pi}_j(\delta_j,\boldsymbol{\delta})}.\label{betaDefn0}
\end{align}
The ISO's problem (\ref{ISO_asymptoticProblemLong}) can then be expressed compactly as
\begin{align}
    \underset{\mathbf{g}\in\mathbb{R}^T,\boldsymbol{\pi}\in\Pi}{\mathrm{Minimize}}\;\;\mathbb{E}_{\boldsymbol{\delta}\sim\mathbb{P}_{\boldsymbol{\theta}}}\bigg[\beta(\boldsymbol{\delta},\mathbf{g},\boldsymbol{\pi})\bigg].\label{ISO_asymptoticProblem}
\end{align}}

{We define three functions based on (\ref{ISO_asymptoticProblem}). First, we define $\mathbf{g}^*:\Theta^{n_s}\to\mathbb{R}^T$ as a function that maps the EV parameters to an energy dispatch sequence that solves (\ref{ISO_asymptoticProblem}). Specifically, for $\psi\in\Theta^{n_s},$ $\mathbf{g}^*(\boldsymbol{\psi})$ denotes an optimal energy dispatch sequence that solves (\ref{ISO_asymptoticProblem}) if the EV parameters are $\boldsymbol{\psi}.$ Similarly, we define $\boldsymbol{\pi}^*:\Theta^{n_s}\to\Pi$ as the function that maps the EVs' parameters to an optimal storage policy that solves (\ref{ISO_asymptoticProblem}). Finally, we define $q^*:\Theta^{n_s}\to\mathbb{R}$ as the function that maps the EVs' parameters to the optimal average cost so that $q^*(\boldsymbol{\psi})$ denotes the optimal value of (\ref{ISO_asymptoticProblem}) if the EV parameters are $\boldsymbol{\psi}.$ We will assume throughout that the cost functions $c^g$ and $c^{s}$ are such that for some $\overline{Q}<\infty$ and for all $\boldsymbol{\psi}\in\Theta^{n_s},$
\begin{align}
    q^*(\boldsymbol{\psi})\leq\overline{Q}.\label{costFinite}
\end{align}}

{Hence, the ISO's objective in the day-ahead market is to compute the functions $\mathbf{g}^*$ and $\boldsymbol{\pi}^*$ at the point $\boldsymbol{\theta}.$ However, if the EVs are strategic, then they may not bid their deadline distributions truthfully in the day-ahead market, and so the ISO has the additional task of eliciting $\boldsymbol{\theta}$ truthfully. We will discuss how the ISO can do this in an ensuing subsection.}

{Note also that if the EVs report their deadline realizations truthfully every day, then the decisions $\mathbf{g}^*(\boldsymbol{\theta})$ and $\boldsymbol{\pi}^*(\boldsymbol{\theta})$, when used on all days, almost surely minimize the time-averaged cost of meeting the demand. That is, the long-term average cost of operating the grid 
\begin{align}
    \limsup_{L\to\infty}\frac{1}{L}\sum_{l=1}^L\bigg[\beta(\boldsymbol{\delta}(l),\mathbf{g}^l,\boldsymbol{\pi}^l)\bigg]\label{ISO_asymptoticProblemBTW}
\end{align}
where $\mathbf{g}^l$ denotes the energy dispatch sequence on day $l$ and $\boldsymbol{\pi}^l$ denotes the storage policy used on day $l,$ is minimized by setting $\mathbf{g}^l=\mathbf{g}^*(\boldsymbol{\theta})$ and $\boldsymbol{\pi}^l=\boldsymbol{\pi}^*(\boldsymbol{\theta})$ for all $l.$ The minimization could be carried out over all adapted policies that determine $\mathbf{g}^l$ and $\boldsymbol{\pi}^l$ on each day $l$.} 


\subsection{The EV Operator's Objective}
On each day $l$, each EV $i$ receives a payment $p_i(l)$ from the ISO in return for leasing its battery to the grid. Denoting by $\mathbf{h}_{i,l}$ the storage sequence of EV $i$ on day $l$, its utility on that day is defined as
\begin{align*}
    u_i(\delta_i(l),\widehat{\delta}_{i}(l),&\mathbf{h}_{i,l})\coloneqq p_i(l)-c_i^{EV}(\delta_i(l),\widehat{\delta}_i(l),\mathbf{h}_{i,l}).
\end{align*}
Each EV $i$'s objective is to maximize its long-term average utility defined as $\liminf_{L\to\infty}\frac{1}{L}\sum_{l=1}^L{u_i(\delta_i(l),\widehat{\delta}_{i}(l),\mathbf{h}_{i,l})}.$  

\subsection{The Market Process}

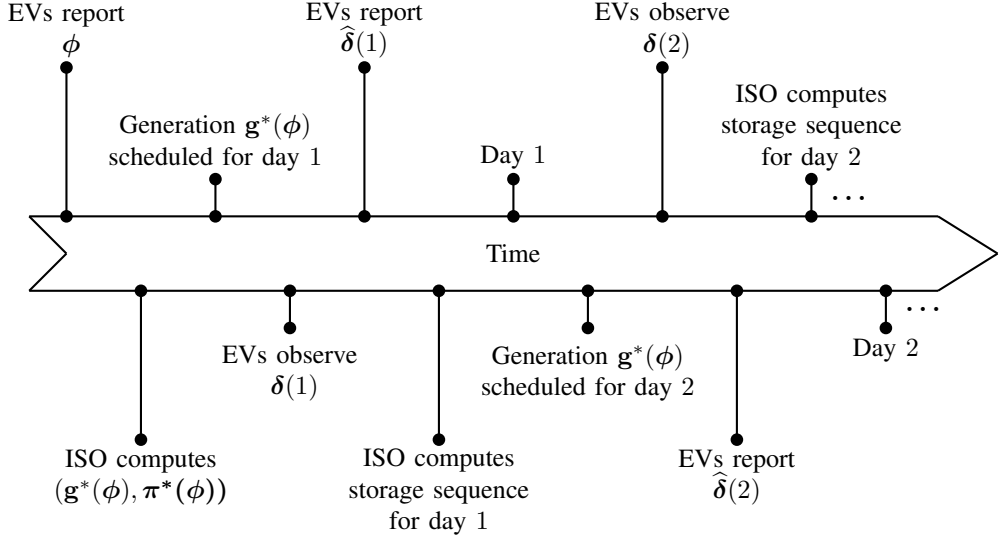
\begin{figure*}
\centering
\begin{adjustbox}{width=1.5\columnwidth}
\begin{tikzpicture}

\draw[black, thick] (1,6) -- (13.2,6);
\draw[black, thick] (1,7) -- (13.2,7);
\draw[black, thick] (13.2,6) -- (14,6.5);
\draw[black, thick] (13.2,7) -- (14,6.5);
\draw[black, thick] (1,6) -- (1.5,6.5);
\draw[black, thick] (1,7) -- (1.5,6.5);


\draw[black, thick, fill] (1.5,7) circle (2pt);
\draw[black, thick] (1.5,7) -- (1.5,9);
\draw[black, thick, fill] (1.5,9) circle (2pt);
\node[align=center] at (1.5,9.5) {\text{EVs report}\\\text{ $\boldsymbol{\phi}$}};

\draw[black, thick, fill] (2.5,6) circle (2pt);
\draw[black, thick] (2.5,6) -- (2.5,4);
\draw[black, thick, fill] (2.5,4) circle (2pt);
\node[align=center] at (2.5,3.5) {\text{ISO computes}\\\text{$(\mathbf{g}^*(\boldsymbol{\phi}),\boldsymbol{\pi^*(\boldsymbol{\phi}))}$}};


\draw[black, thick, fill] (3.5,7) circle (2pt);
\draw[black, thick] (3.5,7) -- (3.5,7.5);
\draw[black, thick, fill] (3.5,7.5) circle (2pt);
\node[align=center] at (3.5,8) {Generation $\mathbf{g}^*(\boldsymbol{\phi})$\\ scheduled for day $1$};

\draw[black, thick, fill] (4.5,6) circle (2pt);
\draw[black, thick] (4.5,6) -- (4.5,5.5);
\draw[black, thick, fill] (4.5,5.5) circle (2pt);
\node[align=center] at (4.5,4.9) {\text{EVs observe}\\\text{ $\boldsymbol{\delta}(1)$}};

\draw[black, thick, fill] (5.5,7) circle (2pt);
\draw[black, thick] (5.5,7) -- (5.5,9);
\draw[black, thick, fill] (5.5,9) circle (2pt);
\node[align=center] at (5.5,9.5) {\text{EVs report}\\\text{$\widehat{\boldsymbol{\delta}}(1)$}};

\draw[black, thick, fill] (6.5,6) circle (2pt);
\draw[black, thick] (6.5,6) -- (6.5,4);
\draw[black, thick, fill] (6.5,4) circle (2pt);
\node[align=center] at (6.5,3.3) {\text{ISO computes}\\\text{storage sequence}\\\text{for day $1$}};


\draw[black, thick, fill] (7.5,7) circle (2pt);
\draw[black, thick] (7.5,7) -- (7.5,7.5);
\draw[black, thick, fill] (7.5,7.5) circle (2pt);
\node[align=center] at (7.5,7.8) {\text{Day $1$}};

\draw[black, thick, fill] (8.5,6) circle (2pt);
\draw[black, thick] (8.5,6) -- (8.5,5.5);
\draw[black, thick, fill] (8.5,5.5) circle (2pt);
\node[align=center] at (8.5,4.9) {Generation $\mathbf{g}^*(\boldsymbol{\phi})$\\ scheduled for day $2$};

\draw[black, thick, fill] (9.5,7) circle (2pt);
\draw[black, thick] (9.5,7) -- (9.5,9);
\draw[black, thick, fill] (9.5,9) circle (2pt);
\node[align=center] at (9.5,9.5) {\text{EVs observe}\\\text{ $\boldsymbol{\delta}(2)$}};

\draw[black, thick, fill] (10.5,6) circle (2pt);
\draw[black, thick] (10.5,6) -- (10.5,4);
\draw[black, thick, fill] (10.5,4) circle (2pt);
\node[align=center] at (10.5,3.5) {\text{EVs report}\\\text{$\widehat{\boldsymbol{\delta}}(2)$}};

\draw[black, thick, fill] (11.5,7) circle (2pt);
\draw[black, thick] (11.5,7) -- (11.5,7.5);
\draw[black, thick, fill] (11.5,7.5) circle (2pt);
\node[align=center] at (11.5,8.2) {\text{ISO computes}\\\text{storage sequence}\\\text{for day $2$}};

\draw[black, thick, fill] (12.5,6) circle (2pt);
\draw[black, thick] (12.5,6) -- (12.5,5.5);
\draw[black, thick, fill] (12.5,5.5) circle (2pt);
\node[align=center] at (12.5,5.2) {\text{Day $2$}};

\node [black] at (12,7.25) {$\boldsymbol{\hdots}$};
\node [black] at (13,5.75) {$\boldsymbol{\hdots}$};

\node[align=center] at (7.5,6.5) {\text{Time}};

\end{tikzpicture}
\end{adjustbox}
\caption{The market chronology: In the day-ahead market before day $1$, each EV $i$ reports a deadline distribution to the ISO, based on which the latter computes the energy dispatch sequence and the storage policy. Then, at the commencement of any day $l$, $l\in\mathbb{Z}_+,$ each EV $i$ observes its deadline $\delta_i(l)$ for that day. Following this, EV $i$ reports $\widehat{\delta}_i(l)$ as its deadline, which could potentially be adapted to $\delta_i^l,\widehat{\delta}_i^{l-1},\boldsymbol{\delta}_{-i}^l,\widehat{\boldsymbol{\delta}}_{-i}^l,\boldsymbol{\theta},$ and $\boldsymbol{\phi}.$ Based on $\widehat{\boldsymbol{\delta}}(l)$, the ISO computes the storage schedule for each EV for that day using the storage policy. Day $l$ then progresses, and the process repeats on day $l+1$.}\label{Fig1}
\end{figure*}

There are two impediments to the ISO operating the grid at the optimal cost $q^*(\boldsymbol{\theta}).$ The first is the ISO's nescience of the parameter vector $\boldsymbol{\theta},$ which renders it incapable of computing the optimal decisions $g^*(\boldsymbol{\theta})$ and $\boldsymbol{\pi}^*(\boldsymbol{\theta})$ in the day-ahead market. As mentioned before, an EV's deadline distribution is its private knowledge and unknown to the ISO. Consequently, the ISO requests each EV to report its parameter in the day-ahead market so that it can compute the optimal energy dispatch sequence and storage policy. However, since the objective of any EV is only to maximize its own utility, it may misreport its parameter if there is a possibility for it to extract a higher utility by doing so than by bidding truthfully. Consequently, we denote by $\phi_i\in\Theta$ the parameter reported by EV $i$ in the day-ahead market, which may or may not be equal to $\theta_i.$ 

Based on the reported parameters $\boldsymbol{\phi}\coloneqq[\phi_1,\hdots,\phi_{n_s}],$ the ISO computes the energy dispatch as $\mathbf{g}^*(\boldsymbol{\phi})$ and the storage policy as $\boldsymbol{\pi}^*(\boldsymbol{\phi})$. As mentioned before, since the deadline distributions of the EVs are assumed to remain the same on all days, it suffices for the EVs to report their parameters just once, and for the ISO to run the day-ahead market just once, namely, before day $1$, to compute the above decisions. Once the ISO computes these quantities, it reuses them on all days.

In the day-ahead market corresponding to any day $l$, the ISO schedules the generator to produce the energy sequence $\mathbf{g}^*(\boldsymbol{\phi})$ on day $l$ and decides on the storage policy $\boldsymbol{\pi}^*(\boldsymbol{\phi})$ for day $l.$ After the day-ahead market closes, the EVs observe their respective deadlines for that day. The ISO requests the EVs to report their deadline realizations at the commencement of day $l$. Based on the reported deadlines, the ISO computes the storage sequence 
of each EV for that day using the policy $\boldsymbol{\pi}^*(\boldsymbol{\phi})$. Being strategic, the EVs may not bid their deadline realizations truthfully, and so we denote by $\widehat{\delta}_i(l)$ the deadline reported by EV $i$ on day $l.$ Having bid $\widehat{\delta}_i(l)$ as its deadline, EV $i$ is obliged to remain connected to the grid until time $\widehat{\delta}_i(l)$ on day $l.$ The entire chronology is illustrated in Fig. \ref{Fig1}.

\subsection{The Mechanism Design Problem}
Having fixed $\mathbf{g}^*(\boldsymbol{\phi})$ and $\boldsymbol{\pi}^*(\boldsymbol{\phi})$ in the day-ahead market, the long-term average utility that EV $i$ accrues is 
\begin{align}
    u_i^\infty(\phi_i&,\widehat{\delta}_i^\infty,\boldsymbol{\phi}_{-i},\widehat{\boldsymbol{\delta}}_{-i}^\infty,\omega^\infty_{\boldsymbol{\pi}^*(\boldsymbol{\phi})})\nonumber\\
    \coloneqq&\liminf_{L\to\infty}\frac{1}{L}\sum_{l=1}^Lu_i\big(\delta_i(l),\widehat{\delta}_i(l),{\boldsymbol{\pi}}^*_i(\widehat{\boldsymbol{\delta}}(l),\omega_{\boldsymbol{\pi}^*(\boldsymbol{\phi})}(l);\boldsymbol{\phi})\big),\label{EViLongTermUtility}
\end{align} 
where $\boldsymbol{\pi}^*(\widehat{\boldsymbol{\delta}}(l),\omega_{\boldsymbol{\pi}^*(\boldsymbol{\phi})}(l);\boldsymbol{\phi})$ denotes the function $\boldsymbol{\pi}_i^*(\boldsymbol{\phi})$ evaluated at $\big(\widehat{\boldsymbol{\delta}}(l),\omega_{\boldsymbol{\pi}^*(\boldsymbol{\phi})}(l)\big).$ Note that in addition to $\phi_i$, the long-term average utility $u_i^\infty$ is also a function of the times $\widehat{\delta}_i^\infty$ that EV $i$ reports. Consequently, if there exists $(\boldsymbol{\phi}_{-i},\widehat{\boldsymbol{\delta}}^\infty_{-i},\omega^\infty_{\boldsymbol{\pi}^*(\boldsymbol{\phi})})$ such that 
EV $i$ obtains a higher value for (\ref{EViLongTermUtility}) by misreporting either or both $\theta_i$ and $\delta_i^\infty$, then it may do so. However, unless all EVs report their respective parameters truthfully in the day-ahead market and report their respective deadlines truthfully ``almost all days," the ISO cannot ensure that the long-term average cost of meeting the demand defined in (\ref{ISO_asymptoticProblemBTW}) approaches the optimal value $q^*(\boldsymbol{\theta})$. 
This brings us to the central problem that is addressed in the paper, namely, that of designing mechanisms that incentivize EVs to report not only their deadline distributions truthfully in the day-ahead market, but also report their deadline realizations truthfully almost all days. Specifically, we aim to design a mechanism that renders truth-telling a dominant strategy so that for every EV $i$, its average utility $u_i^\infty(\phi_i,\widehat{\delta}_i^\infty,\boldsymbol{\phi}_{-i},\widehat{\boldsymbol{\delta}}_{-i}^\infty,\omega^\infty_{\boldsymbol{\pi}^*(\boldsymbol{\phi})})$ is maximized by setting $\phi_i=\theta_i$ and $\widehat{\delta}_i(l)=\delta_i(l)$ for all $l\in\mathbb{Z}_+,$ regardless of what $(\boldsymbol{\phi}_{-i},\widehat{\boldsymbol{\delta}}_{-i}^\infty,\omega^\infty_{\boldsymbol{\pi}^*(\boldsymbol{\phi})})$ is. The next section develops the mechanism and establishes the incentive and optimality properties guaranteed by it.

\section{Mechanism for Trading Storage Capacity of EVs with Stochastic Deadlines}\label{Section_Market}
In this section, we develop a mechanism that renders truth-telling an individually rational dominant strategy for every EV. 

First, the ISO computes $\mathbf{g}^*(\boldsymbol{\phi})$ and $\boldsymbol{\pi}^*(\boldsymbol{\phi})$ in the day-ahead market as the generator's energy dispatch sequence and the storage policy. Recall that these quantities solve the stochastic program
\begin{align}
    \underset{\mathbf{g},\boldsymbol{\pi}}{\mathrm{Min.}}\;\; c^g(\mathbf{g})+\mathbb{E}_{\boldsymbol{\delta}\sim\mathbb{P}_{\boldsymbol{\phi}}}\Big[\widehat{c^s}(\boldsymbol{\delta},\mathbf{g},\boldsymbol{\pi})-\sum_{j=1}^{n_s}\widehat{\pi}_j(\delta_j,\boldsymbol{\delta})\Big],\label{ISO_OneShot}
\end{align}
and note that the solutions of (\ref{ISO_asymptoticProblem}) and (\ref{ISO_OneShot}) coincide if $\boldsymbol{\phi}=\boldsymbol{\theta}.$ 

We now describe the payment rule. The payment rule consists of each EV receiving two payments on each day, namely, a ``day-ahead payment" that is determined based on the parameters reported to the ISO in the day-ahead market, and an ``end-of-the-day settlement" that is determined at the end of each day based on a comparison of the the actual departure profile of the EVs with the deadline distributions reported in the day-ahead market. 

\subsection{The Day-Ahead Payment}
The day-ahead payment takes the form of a VCG payment. For each $i\in\{1,\hdots,n_s\},$ define $\mathbb{P}_{\boldsymbol{\phi}_{-i}}\coloneqq\mathbb{P}_{\phi_1}\times\hdots\times\mathbb{P}_{\phi_{i-1}}\times\mathbb{P}_{\phi_{i+1}}\times\hdots\times\mathbb{P}_{\phi_{n_s}}$ and let $q^*(\boldsymbol{\phi}_{-i})$ be the optimal value of the stochastic program
\begin{align*}
    \underset{\mathbf{g},\boldsymbol{\pi}}{\mathrm{Min.}}\;\; c^g(\mathbf{g})+\mathbb{E}_{\boldsymbol{\delta}_{-i}\sim\mathbb{P}_{\boldsymbol{\phi}_{-i}}}\Big[\widehat{c^s}(\boldsymbol{\delta_{-i}},\mathbf{g},\boldsymbol{\pi})
    -\sum_{j\neq i}\widehat{\pi}_j(\delta_j,\boldsymbol{\delta}_{-i})\Big].
\end{align*}
Note that this is the optimization problem that the ISO would have had to solve in the day-ahead market if EV $i$ were absent from the system. 
The day-ahead payment of EV $i$, $i\in\{1,\hdots,n_s\},$ is defined as 
\begin{align}
    p_i^{DA}(\phi_i&,\boldsymbol{\phi}_{-i})\coloneqq q^*(\boldsymbol{\phi}_{-i})
    -\Bigg[ c^g(\mathbf{g}^{*}(\boldsymbol{\phi}))\nonumber\\
    +&\mathbb{E}_{\boldsymbol{\delta}\sim\mathbb{P}_{\boldsymbol{\phi}}}\Big\{\widehat{c^s}(\boldsymbol{\delta},\mathbf{g}^*(\boldsymbol{\phi}),\boldsymbol{\pi}^*(\boldsymbol{\phi}))-\sum_{j\neq i}\widehat{\pi}_j^*(\delta_j,\boldsymbol{\delta};\boldsymbol{\phi})\Big\}  \Bigg].\label{POE}
\end{align}



\subsection{The End-of-the-Day Settlement}
In addition to the day-ahead payment, the ISO also makes an ``end-of-the-day settlement" to each EV $i$ on each day $l$. 
A negative value of this quantity indicates a ``penalty" that should be paid by the EV to the ISO. 

For $i\in\{1,\hdots,n_s\}$, $t\in\{1,\hdots,T\}$, and $l\in\mathbb{Z}_+,$ define 
\begin{align}
    {f}_{i,t}(l,\widehat{\delta}_i^l,\phi_i)\coloneqq\bigg[\frac{1}{l}\sum_{l'=1}^{l}\mathds{1}_{\{\widehat{\delta}_i(l')=t\}}\bigg]-\mathbb{P}_{\phi_i}(t)\label{fdefn}
\end{align}
where $\mathbb{P}_{\phi_i}(t)$, recall, denotes the probability that a random variable distributed according to $\mathbb{P}_{\phi_i}$ takes the value $t.$ 
Let $\{r\}$ be any nonnegative sequence such that for some $L_r\in\mathbb{N}$ and some $\gamma>\frac{1}{2},$
\begin{align}
    {r(l)}\geq\sqrt{\frac{\ln{l^{\gamma}}}{l}}\label{romega}
\end{align}
for all $l\geq L_r,$ and
\begin{align}
    \lim_{l\to\infty}r(l)=0.\label{rgoestozero}
\end{align}
For all $l\in\mathbb{Z}_+$ and all $i\in\{1,\hdots,n_s\}$, define event
\begin{align}
    E_i(l,\widehat{\delta_i^l},\phi_i)\coloneqq\{\max_{t\in\{1,\hdots,T\}}\vert{f}_{i,t}(l,\widehat{\delta}_i^l,\phi_i)\vert\geq r(l)\}.\label{Edefn}
\end{align}
Let $\{J_p\}$ be any nonnegative sequence such that
\begin{align}
    \lim_{l\to\infty}\frac{J_p(l)}{l}=\infty.\label{JpOmegal}
\end{align}

The end-of-the-day settlement of EV $i$ on day $l$ is defined as 
\begin{align}
    p_i^{S}(l,\boldsymbol{\phi},\widehat{\delta}^l_i,\widehat{\boldsymbol{\delta}}(l),\omega_{\boldsymbol{\pi}^*(\boldsymbol{\phi})}(l))\;\;\;\;\;\;\;\;\;\;\;\;\;\;\;\;\;&\nonumber\\
    \coloneqq\bigg[\mathbb{E}_{\boldsymbol{\delta}\sim\mathbb{P}_{\boldsymbol{\phi}}}\{\widehat{\pi}_i^*(\delta_i,\boldsymbol{\delta};\boldsymbol{\phi})\}-{\pi}_i^*(\widehat{\delta}_i(l),&\widehat{\boldsymbol{\delta}}(l),\omega_{\boldsymbol{\pi}^*(\boldsymbol{\phi})}(l);\boldsymbol{\phi})\bigg]\nonumber\\
    &-J_p(l)\mathds{1}_{\{E_i(l,\widehat{\delta}_i^l,\phi_i)\}}.\label{piRT}
\end{align}
The term 
within the square brackets in the above RHS has the interpretation as the difference between the expected amount of energy injected into EV $i$ on day $l$ as determined in the day-ahead market and the actual energy injected into EV $i$ on day $l$. Note that if all EVs are truthful, then the long-term average of this quantity approaches zero almost surely. The second component of the end-of-the-day settlement essentially penalizes EV $i$ for discrepancies between the empirical distribution of its reported deadlines and the distribution $\mathbb{P}_{\phi_i}$ that it reports in the day-ahead market. 

The total payment $p_i$ received by EV $i$ on day $l$ is the sum of its day-ahead payment and its end-of-the-day settlement:
\begin{align}
    p_i(l&,\phi_i,\boldsymbol{\phi}_{-i},\widehat{\delta}_i^l,\widehat{\boldsymbol{\delta}}(l),\omega_{\boldsymbol{\pi}^*(\boldsymbol{\phi})}(l))\nonumber\\
    &=p_i^{DA}(\phi_i,\phi_{-i})+p_i^{S}(l,\boldsymbol{\phi},\widehat{\delta}_i^l,\widehat{\boldsymbol{\delta}}(l),\omega_{\boldsymbol{\pi}^*(\boldsymbol{\phi})}(l)).\label{payment}
\end{align}

{Theorem 1 establishes the incentive and efficiency properties of the mechanism defined by the decision rule (\ref{ISO_OneShot}) and the payment rule (\ref{payment}). Before presenting the theorem, we explain at a high level why the proposed mechanism provides the desired incentive properties.}

{Arbitrarily fix an EV $i$. As mentioned before, one of the functionalities of the end-of-the-day settlement is to penalize the EV for deviations of its empirically observed departure times from $\mathbb{P}_{\phi_i}$. To enforce this, on each day $l$ and for each $t\in\{1,\hdots,T\},$ the end-of-the-day settlement function constructs a window of size $r(l)$ centered at $\mathbb{P}_{\phi_i}(t)$ and penalizes the EV if the empirical frequency $\frac{1}{l}\sum_{l'=1}^l\mathds{1}_{\{\widehat{\delta}_i(l')=t\}}$ falls outside the window. }

{Now, the window size must be designed carefully so as to balance two competing objectives. On the one hand, the window size must approach zero as $l$ tends to infinity. If not, the set of sequences from which the EV can choose its real-time bids $\widehat{\delta}_i^\infty$ without incurring a penalty would be ``large," thereby resulting in the violation of incentive compatibility. On the other hand, if the window size shrinks too quickly, then the aforementioned empirical frequency sequence of truthful bidders will fall outside the window infinitely often. This would result in truthful bidders paying a penalty infinitely often, thereby resulting in the violation of their individual rationality. This brings us to the question of what the appropriate rate is at which the window size should decay. Condition (\ref{romega}) answers this question. To provide some intuition for this condition, note that the empirical frequency $\frac{1}{l}\sum_{l'=1}^l\mathds{1}_{\{\delta_i(l')=t\}}$ of EV $i$'s true deadlines is approximately normally distributed for large $l$ with a standard deviation that scales as ${\frac{1}{\sqrt{l}}}$. Hence, scaling the window size also at the same rate would result in the probability of empirical frequencies of truthful bids falling outside the window to remain at some fixed value which does not scale with $l$. To avoid this, the window size must scale slower than at least ${\frac{1}{\sqrt{l}}}.$ Lemma \ref{Lemma_honesty} shows that by scaling it only ``slightly" slower than this rate, namely, at the rate specified by (\ref{romega}), the truthful bidders are guaranteed to not incur a penalty.}

{Once an appropriate window sequence is chosen, EV $i$ becomes subject to the constraint that its real-time bid sequence $\widehat{\delta}_i^\infty$ has a histogram that ``looks like" $\mathbb{P}_{\phi_i}.$ If not, the average penalty that it would pay will be infinite -- a property that stems from (\ref{JpOmegal}) and is established in Lemma \ref{lemmaInfinitePenalty}. We refer to this constraint as C1. Henceforth, it suffices to restrict attention only to those strategies that satisfy C1.}

{For $\theta\in\Theta,$ denote by $F_{\theta}$ the cumulative distribution function corresponding to parameter ${\theta}$. Suppose first that $\phi_i$ is such that $F_{\theta_i}(t')>F_{\phi_i}(t')$ for some $t'\in\{1,\hdots,T\},$ and suppose that the EV reports $\widehat{\delta}_i^\infty$ such that C1 is satisfied. Then, it is easy to see that regardless of how $\widehat{\delta}_i^\infty$ is fabricated, it will be larger than the EV's true deadline $\delta_i^\infty$ at least $F_{\theta_i}(t')-F_{\phi_i}(t')$ fraction of the days on average. This is established in Lemma \ref{LemmaLipschitz}. Hence, letting $\alpha(\theta_i,\phi_i)\coloneqq\sup_{t\in\{1,\hdots,T\}}F_{\theta_i}(t)-F_{\phi_i}(t),$ EV $i$ must miss its deadline at least $\alpha(\theta_i,\phi_i)$ fraction of the days on average, thereby incurring a cost of at least $J_m\alpha(\theta_i,\phi_i)$ on average for missing its deadlines. What about the payment that it receives though as a result of bidding such a $\phi_i$? Is there a possibility for the EV to obtain a higher payment by bidding such a $\phi_i$ than by bidding $\theta_i$? The answer is in the affirmative. To elaborate, by remaining connected to the grid beyond its deadline, the EV could aid the ISO in reducing the cost of meeting the demand. The VCG payment rule would reflect this positive externality and so the EV could potentially obtain a larger payment every day in the day-ahead market by bidding $\phi_i$ than by bidding $\theta_i$. Now, is the increase in payment sufficient to outweigh the cost incurred for missing deadlines? The answer is in the negative, as established in Theorem 1. Basically, the cost $q^*$ possesses a Lipschitz-like property in that the reduction in the ISO's cost due false bid $\phi_i$ is bounded by a term proportional to $\alpha(\theta_i,\phi_i).$ This is established in Lemma \ref{LemmaLipschitz}. The composite payment rule is designed so that the EV's utility function inherits this property, and it follows from this that the increase in the payment received by the EV due to a false bid is upper bounded by a quantity that is insufficient to make up for the cost incurred by missing deadlines. }

{This leaves the EV with just one option for misreporting in the day-ahead market, namely, bidding $\phi_i$ such that $F_{\phi_i}(t)\geq F_{\theta_i}(t)$ for all $t\in\{1,\hdots,T\}.$ If this condition holds and $\phi_i\neq\theta_i$, we say that $\phi_i\prec\theta_i$, and term such a $\phi_i$ as an ``underbid" for $\theta_i$. Now, for every $\boldsymbol{\phi}_{-i},$ the function $q^*(\;\cdot\;,\;\boldsymbol{\phi}_{-i})$ is monotonic in that if $\phi_i\prec\theta_i$, then $q^*(\phi_i,\boldsymbol{\phi}_{-i})\geq q^*(\theta_i,\boldsymbol{\phi}_{-i}).$ This is established in Lemma \ref{monotonic}. The VCG payment rule would reflect this negative externality and so the EV by underbidding $\theta_i$ would obtain a lower payment every day in the day-ahead market than by bidding $\theta_i$ truthfully. What about the cost that it incurs though as a result of underbidding? Is there a possibility for it to incur a lower cost by underbidding $\theta_i$ than by bidding it truthfully? The answer is in the affirmative. Underbidding in the day-ahead market affords the EV an extensive strategy space in real time using which it can misreport its deadline realizations without incurring a penalty or missing any deadlines. 
The EV may be able to exploit this flexibility to lower its average cost. This is best illustrated with the help of an example. Suppose that an EV has a deterministic deadline $\tau,$ where $\tau\in\{1,\hdots,T\}.$ Instead of bidding this truthfully, the EV could bid any distribution that has support $\{1,\hdots,\tau\}.$ Consequently, on any given day, the EV can choose any value in this set to disconnect from the grid, and every one of these choices would ensure that it doesn't miss its deadline. Hence, only C1 remains to be satisfied, and this constraint is relatively easy to satisfy if the EV on each day adapts its reported deadline to all deadlines reported in the past. This still leaves the EV with ample freedom to choose its real time bids, which it could possibly exploit to disconnect with a higher level of charge, or equivalently with a lower cost, than the cost it would incur if it bids truthfully. Now, is the cost reduction sufficient to outweigh the reduced payments? The answer in the negative, as established in Theorem 1. Basically, the first term of the end-of-the-day settlement (\ref{piRT}) compares the battery charge with which EV $i$ disconnects on day $l$ with that expected in the day-ahead market. Any bias in this quantity appears as a debit to the EV at the end of the day. It follows from this that the cost reduction attained via strategic real-time bidding is insufficient to outweigh the reduced day-ahead payments. }

{In summary, the EV can neither report $\phi_i$ such that $\phi_i\prec\theta_i$ nor $\phi_i$ such that $\alpha(\theta_i,\phi_i)>0.$ The only remaining option is for it to report $\phi_i=\theta_i.$ Theorem 1 establishes this rigorously.} 


{We now present the theorem. The first statement of the theorem states that truthful bidding in both the day-ahead market and in real time is a dominant strategy for every EV. The second statement states that under a certain mild condition, truthful bidding is not only a dominant strategy but it is also the unique dominant strategy in a certain sense. Specifically, under a certain mild condition, an EV bidding its deadline distribution truthfully in the day-ahead market and bidding its deadline realization truthfully ``almost all days" is the {unique} dominant strategy. The third statement states that truthful bidding is individually rational for every EV regardless of how the other EVs bid. The final statement shows that the mechanism aligns the EVs' objectives with the ISO's objective in that every EV employing a selfish utility-maximizing strategy automatically results in the ISO satisfying the demand at minimum possible cost.}

\begin{theorem}
Suppose that the ISO determines $\mathbf{g}^{*}(\boldsymbol{\phi})$ and $\boldsymbol{\pi}^{*}(\boldsymbol{\phi})$ as a solution to (\ref{ISO_OneShot}) and determines the payments according to (\ref{payment}). Then, for $J_m$ sufficiently large, the following hold.
\begin{enumerate}
    \item For every $i\in\{1,\hdots,n_s\}$ and every $\theta_i\in\Theta,$ there exists $\mathcal{E}_i\subset\{1,\hdots,T\}^\infty$ with $\mathbb{P}_{\theta_i}^\infty(\mathcal{E}_i)=0$ such that for every $\delta_i^\infty\notin\mathcal{E}_i,$
\begin{align}
    u_i^\infty(\theta_i,{\delta}_i^\infty,&{\boldsymbol{\phi}}_{-i},\widehat{\boldsymbol{\delta}}_{-i}^\infty,\omega^\infty_{\boldsymbol{\pi}^*(\theta_i,\boldsymbol{\phi}_{-i})})\nonumber\\
    &\geq u_i^\infty(\phi_i,\widehat{\delta}_i^\infty,{\boldsymbol{\phi}}_{-i},\widehat{\boldsymbol{\delta}}_{-i}^\infty,\omega^\infty_{\boldsymbol{\pi}^*(\boldsymbol{\phi})}) \label{DSIC}
\end{align}
for every $\boldsymbol{\phi}$, $\widehat{\boldsymbol{\delta}}^\infty$, $\omega^\infty_{\boldsymbol{\pi}^*(\theta_i,\boldsymbol{\phi}_{-i})},$ and $\omega^\infty_{\boldsymbol{\pi}^*(\boldsymbol{\phi})}$. 

I.e., for every EV $i$, truth-telling is $\mathbb{P}_{\theta_i}^\infty-$ almost surely a dominant strategy.

\item Let $i\in\{1,\hdots,n_s\}$ and suppose that $\theta_i$ is such that for all $\phi_i\neq\theta_i,$
\begin{align}
    q^*(\phi_i,\boldsymbol{\phi}_{-i})\neq q^*(\theta_i,\boldsymbol{\phi}_{-i})\label{uniqueDScondition}
\end{align}
for some $\boldsymbol{\phi}_{-i}\in\Theta^{n_s-1}.$ If for some $\phi_i\in\Theta$, there exists $\mathcal{E}_i\subset\{1,\hdots,T\}^\infty$ with $\mathbb{P}_{\theta_i}^\infty(\mathcal{E}_i)=0$ such that for every $\delta_i^\infty\notin\mathcal{E}_i,$ $\boldsymbol{\phi}_{-i},$ $\widehat{\boldsymbol{\delta}}_{-i}^\infty,$ $\omega^\infty_{\boldsymbol{\pi}^*(\boldsymbol{\phi})},$ and $\omega^\infty_{\boldsymbol{\pi}^*(\theta_i,\boldsymbol{\phi}_{-i})},$ there exists $\widehat{\delta}_i^\infty$ such that
\begin{align}
    u_i^\infty(\phi_i,&\widehat{\delta}_i^\infty,{\boldsymbol{\phi}}_{-i},\widehat{\boldsymbol{\delta}}_{-i}^\infty,\omega^\infty_{\boldsymbol{\pi}^*(\boldsymbol{\phi})})\nonumber\\
    &=u_i^\infty(\theta_i,{\delta}_i^\infty,{\boldsymbol{\phi}}_{-i},\widehat{\boldsymbol{\delta}}_{-i}^\infty,\omega^\infty_{\boldsymbol{\pi}^*(\theta_i\boldsymbol{\phi}_{-i})}),\label{equalUtilities}
\end{align}
then, 
\begin{align}
    \phi_i=\theta_i,\label{thetaTruthful}
\end{align}
and 
\begin{align}
    \lim_{L\to\infty}\frac{1}{L}\sum_{l=1}^L \mathds{1}_{\{\widehat{\delta}_i(l)\neq\delta_i(l)\}}=0.\label{deltaTruthful}
\end{align}
I.e., for every $i\in\{1,\hdots,n_s\}$ such that (\ref{uniqueDScondition}) holds, truth-telling in the day-ahead market and truth-telling on almost all days is $\mathbb{P}_{\theta_i}^\infty-$ almost surely the unique dominant strategy for EV $i$.
\item For every $i\in\{1,\hdots,n_s\}$ and every $\theta_i\in\Theta,$ there exists $\mathcal{E}_i\subset\{1,\hdots,T\}^\infty$ with $\mathbb{P}_{\theta_i}^\infty(\mathcal{E}_i)=0$ such that for all $\delta_i^\infty\notin\mathcal{E}_i,$ 
\begin{align}
    u_i^\infty(\theta_i,{\delta}_i^\infty,{\boldsymbol{\phi}}_{-i},\widehat{\boldsymbol{\delta}}_{-i}^\infty,\omega^\infty_{\boldsymbol{\pi}^*(\theta_i,\boldsymbol{\phi}_{-i})})\geq 0\label{IR}
\end{align}
for all $\boldsymbol{\phi}_{-i},$ $\widehat{\boldsymbol{\delta}}_{-i}^\infty,$ and $\omega^\infty_{\boldsymbol{\pi}^*(\theta_i,\boldsymbol{\phi}_{-i})}.$

I.e., for every EV $i,$ truth-telling is $\mathbb{P}_{\theta_i}^\infty-$ almost surely individually rational. 
\item If (\ref{uniqueDScondition}) and (\ref{equalUtilities}) hold for every $i\in\{1,\hdots,n_s\},$ then 
\begin{align}
    \limsup_{L\to\infty}\frac{1}{L}\sum_{l=1}^{L}\beta\big(\widehat{\boldsymbol{\delta}}(l),\mathbf{g}^*(\boldsymbol{\phi}),\boldsymbol{\pi}^*(\boldsymbol{\phi})\big)=q^*(\boldsymbol{\theta})\label{efficiencyOfMarket}
\end{align}
$\mathbb{P}_{\boldsymbol{\theta}}^\infty-$ almost surely.

I.e., if every EV employs a dominant strategy, then the time-averaged cost at which the ISO satisfies the demand is equal to the optimal average cost at which it satisfies the demand if all EVs are truthful.
\end{enumerate}
\end{theorem}
\begin{proof}
See Appendix \ref{TheoremProof}. 
\end{proof}

\section{Algorithms and Numerical Results}\label{numericalResults}

Implementing the mechanism presented in the previous section and operating the grid optimally requires the ISO to solve the stochastic program (\ref{ISO_OneShot}) in the day-ahead market. Note that (\ref{ISO_OneShot}) is essentially an economic dispatch problem for a power system that has EV storage integrated into it. One of the primary difficulties in solving (\ref{ISO_OneShot}) arises from the fact that its decision space is infinite-dimensional; the optimization algorithm must search over the space of all implementable storage policies in order to solve (\ref{ISO_OneShot}). In what follows, we show that for an important special case which the ISO frequently encounters in practice, (\ref{ISO_OneShot}) can be reduced to a finite-dimensional optimization problem. The special case corresponds to the situation where the production function $c^s$ of the reserves is additively separable over time. In today's electricity markets, not only is the function $c^s$ additively separable, it is also linear, and so is the function $c^g$. 
We first present the approach by which (\ref{ISO_OneShot}) reduces to a finite-dimensional optimization problem, and then use it in our simulations to compute the expected cost savings.

A production function $c^s:\mathbb{R}^T\to\mathbb{R}$ is said to be \emph{additively separable over time} if there exist functions $c^s_1,\hdots,c^s_T:\mathbb{R}\to\mathbb{R}$ such that
\begin{align}
    c^s([g_s(1),\hdots,g_s(T)])=\sum_{t=1}^Tc^s_t(g_s(t))\label{csForm}
\end{align}
for all $\mathbf{g}_s\in[G^s_{\mathrm{min}},G^s_\mathrm{max}]^T,$ where $G^s_\mathrm{min}$ and $G^s_\mathrm{max}$ denote the minimum and maximum generation capacity of the reserves. 
Throughout this section, we assume that $c^s$ is of the form (\ref{csForm}).

The reduction technique stems from the rather simple observation that the stochastic program (\ref{ISO_OneShot}) can equivalently be expressed as
\begin{align}
    \underset{\mathbf{g}}{\mathrm{Min.}}\bigg\{\underset{\boldsymbol{\pi}}{\mathrm{Min.}}\;\; \mathbb{E}_{\boldsymbol{\delta}\sim\mathbb{P}_{\boldsymbol{\theta}}}\big[\beta(\boldsymbol{\delta},\mathbf{g},\boldsymbol{\pi})\big]\bigg\},\label{twoStage}
\end{align}
which suggests a two-stage approach to solve the problem. Specifically, if the ``inner program" $$\underset{\boldsymbol{\pi}}{\mathrm{Min.}}\;\; \mathbb{E}_{\boldsymbol{\delta}\sim\mathbb{P}_{\boldsymbol{\theta}}}\big[\beta(\boldsymbol{\delta},\mathbf{g},\boldsymbol{\pi})\big],\;\; $$ can be solved for any $\mathbf{g}$ to result in a storage policy $\boldsymbol{\pi}^\dagger(\mathbf{g})$, then (\ref{twoStage}) can be solved by solving the ``outer program"
\begin{align}
    \underset{\mathbf{g}}{\mathrm{Min.}}\;\;\overline{\beta}(\mathbf{g}),\label{outerProgram}
\end{align}
where $\overline{\beta}(\mathbf{g})\coloneqq\mathbb{E}_{\boldsymbol{\delta}\sim\mathbb{P}_{\boldsymbol{\theta}}}\big[\beta(\boldsymbol{\delta},\mathbf{g},\boldsymbol{\pi}^\dagger(\mathbf{g}))\big].$ This outer program is finite-dimensional. While it may not be convex or otherwise solvable by polynomial-time algorithms, the fact that the ISO has to solve it just once, and that it consists of only few tens of decision variables in practice, implies that its complexity may not be of significant concern. 

The above reduction technique hinges on the ability to solve the inner program which is infinite dimensional. For the case when $c^s$ is additively separable, the inner program can be formulated as a finite-horizon Markov Decision Process (MDP), thereby rendering it solvable via dynamic programming. In what follows, we define the state space, the action set, the transition kernels, the stage costs, and the terminal costs which define the MDP.

\subsection{States}

On any day $l,$ we define the state $s^{EV}_i(t)$ of EV $i$ at  the end of time interval $t$ as $s^{EV}_i(t)\coloneqq(e_i(t),h_i(t))$ where $e_i(t)$ is a $0-1$ variable indicating whether or not EV $i$ is connected to the grid at the end of time interval $t$, and $h_i(t)$ is the amount of energy stored in EV $i$ at the end of time interval $t.$ If the EV is not connected to the grid at the end of time interval $t,$ then $h_i(t)$ is defined as the amount of energy that was stored in the EV at the time that it disconnected from the grid. 

We denote by $\mathcal{H}_i$ the set of energy levels to which EV $i$'s battery can be charged. While typically $\mathcal{H}_i=[0,B_i]$, we allow for $\mathcal{H}_i$ to be more general, namely, we allow for it to be any subset of $[0,B_i].$ 

Consequently, the state space $\mathcal{S}^{EV}_i$ of EV $i$ is  $$\mathcal{S}^{EV}_i\coloneqq\{0,1\}\times\mathcal{H}_i.$$  The initial state $s^{EV}_i(0)$ of every EV $i$ is $(1,0),$ that is, every EV is connected to the grid at the beginning of time interval $1$, and no energy has been stored in any EV by the ISO at that time.

Based on the state of each EV, we define the state $\mathbf{s}(t)$ of the entire EV-integrated power system as $$\mathbf{s}(t)\coloneqq(s^{EV}_1(t),\hdots,s^{EV}_{n_s}(t)).$$ Consequently, the state space $\mathcal{S}$ of the system is
\begin{align*}
    \mathcal{S}=\mathcal{S}^{EV}_1\times\hdots\times\mathcal{S}^{EV}_{n_s},
\end{align*}
and its initial state is $\mathbf{s}(0)=\big((1,0),\hdots,(1,0)\big).$

\subsection{Actions}

For $i\in\{1,\hdots,n_s\}$ and $t\in\{1,\hdots,T\},$ we define the \emph{EV-action} for EV $i$ at time $t$ as the energy storage decision that the ISO makes for EV $i$ at time interval $t,$ namely, the amount of energy that the ISO decides to inject into EV $i$ at time $t.$ We denote this by $a_i(t),$ and a negative value for $a_i(t)$ denotes energy discharge. We define the \emph{EV-level action set} $\mathcal{A}^{EV}_i$ of EV $i$ as 
\begin{align*}
    \mathcal{A}^{EV}_i=[-B_i,B_i],
\end{align*}
which contains every action that the ISO can ever take for EV $i$.

Now, The constraints that (i) the ISO cannot discharge from an EV more energy than it has stored in the EV, (ii) the ISO cannot exceed the EV's battery capacity, (iii) the ISO cannot charge or discharge energy from the EV if it is no longer connected to the grid, and (iv) the amount of energy stored in any EV $i$ must belong to $\mathcal{H}_i$, translate into constraints on what actions from the set $[-B_i,B_i]$ the ISO can take when EV $i$ is in a given state. This results in the EV-level action sets being state-dependent. Specifically, for $s^{EV}_i=(e_i,h_i)\in\mathcal{S}^{EV}_i,$ the set $\mathcal{A}^{EV}_{s^{EV}_i}$ of feasible actions that the ISO can take when EV $i$ is in state ${s}^{EV}_i$ is 
\begin{align}
    \mathcal{A}^{EV}_{s^{EV}_i}\coloneqq\begin{cases}\{a\in\mathcal{A}^{EV}_i:a+h_i\in\mathcal{H}_i\}&;\;\;\mathrm{if}\;\;e_i=1,\nonumber\\
    \;\;\;\;\;\;\;\;\{0\}&;\;\;\mathrm{if}\;\;e_i=0.
    \end{cases}
\end{align}

Based on the sets $\mathcal{A}^{EV}_1,\hdots,\mathcal{A}^{EV}_{n_s}$, we define the system-level action set $\mathcal{A}$, or simply the action set, as 
\begin{align*}
    \mathcal{A}\coloneqq\mathcal{A}^{EV}_1\times\hdots\times\mathcal{A}^{EV}_{n_s}.
\end{align*}
The set $\mathcal{A}$ collects every combination of actions that the ISO can ever take for all EVs.


Since not all actions in the EV-level action sets are feasible for all EV states, not all actions in the action set are feasible for all system states. That is, the feasible action sets are state-dependent. It is easy to verify that for $\mathbf{s}\in\mathcal{S},$ the set $\mathcal{A}_\mathbf{s}$ of feasible actions that the ISO can take when the system is in state $\mathbf{s}$ is 
\begin{align*}
    \mathcal{A}_\mathbf{s}=\mathcal{A}^{EV}_{s_1}\times\hdots\times\mathcal{A}^{EV}_{s_{n_s}},
\end{align*}
where $s_i$ denotes the $i^{th}$ element of $\mathbf{s}.$


\subsection{Transition kernels}\label{subsect_transitionKernel}
Suppose that EV $i$'s state at the end of time interval $t$ is $(1,h_i(t)),$ and that the ISO takes a feasible action $\mathbf{a}$ in time interval $t+1.$ Then, the only states that the EV can transit to at the end of time interval $t+1$ are $(0,h_i(t)+a_i)$ and $(1,h_i(t)+a_i).$ The probability that it transits to the state $(0,h_i(t)+a_i)$ is the conditional probability that EV $i$ disconnects from the grid at time $t+1$ given that it has remained connected until time $t$, which is equal to $\frac{\mathbb{P}_{\theta_i}(t+1)}{1-F_{\theta_i}(t)}$. The probability that it transits to the state $(1,h_i(t)+a_i)$ is $1-\frac{\mathbb{P}_{\theta_i}(t+1)}{1-F_{\theta_i}(t)}$.

On the other hand, if EV $i$ is not connected to the grid at time $t$ so that its state at time $t$ is $(0,h_i(t))$ for some $h_i(t)\in\mathcal{H}_i,$ then, regardless of what feasible action the ISO takes, the only state that the EV can transit to at time $t+1$ is $(0,h_i(t))$, which it does with probability $1$. This yields the EV-level probability transition law for EV $i$ as

\begin{footnotesize}
\begin{align*}
    P^{EV}_i&\big(s^{EV}_i(t+1)\big\vert s^{EV}_i(t),a_i(t+1)\big)\nonumber\\
    =&\begin{cases}
    \frac{\mathbb{P}_{\theta_i}(t+1)}{1-F_{\theta_i}(t)}\mathds{1}_{h_i(t+1)=h_i(t)+a_i(t+1)}&{;}\;\begin{bmatrix}e_i(t)\\e_i(t+1)\end{bmatrix}=\begin{bmatrix}1\\0\end{bmatrix}\\
    [1-\frac{\mathbb{P}_{\theta_i}(t+1)}{1-F_{\theta_i}(t)}]\mathds{1}_{h_i(t+1)=h_i(t)+a_i(t+1)}&{;}\;\begin{bmatrix}e_i(t)\\e_i(t+1)\end{bmatrix}=\begin{bmatrix}1\\1\end{bmatrix}\\
    \mathds{1}_{\{s^{EV}_i(t+1)=s^{EV}_i(t)\}}&{;}\;\;\;\;e_i(t)=0.
    \end{cases}
\end{align*}
\end{footnotesize}

The transition kernel for the entire system can be computed based on the above EV-level probability transition laws. Specifically, for $t\in\{0,\hdots,T-1\},$ if the system is in state $\mathbf{s}$ at the end of time interval $t$ and the ISO takes action $\mathbf{a}\in\mathcal{A}_{\mathbf{s}}$ at time interval $t+1,$ then the probability that it transitions to state $\mathbf{s}'$ at the end of time interval $t+1$ is 
\begin{align}
    P_{t+1}(\mathbf{s}'\vert\mathbf{s},\mathbf{a})\;\;\;\;\;\;\;\;\;\;\;\;\;\;\;\;\;\;\;\;\;\;\;\;\;\;\;\;&\nonumber\\
    \coloneqq\mathrm{Pr}(\mathbf{s}(t+1)=\mathbf{s}'\vert\mathbf{s}(t)=\mathbf{s},&\mathbf{a}(t+1)=\mathbf{a})\nonumber\\
    &=\prod_{i=1}^{n_s}P_i^{EV}(s'_i\big\vert s_i,a_i),
\end{align}
where $s_i',s_i,$ and $a_i$ denote the $i^{th}$ component of $\mathbf{s}',\mathbf{s}$ and $\mathbf{a}$ respectively. 

\subsection{Stage costs}
The stage costs capture the costs incurred by the reserves at each time interval. 
For $t\in\{0,\hdots,T-1\},$ the stage cost $c_{t+1}(\mathbf{s}(t),\mathbf{a}(t+1))$ is defined as the cost incurred by the reserves at time interval $t+1$ if action $\mathbf{a}(t+1)$ is taken at that time. The demand-supply mismatch at time interval $t+1$ if action $\mathbf{a}(t+1)$ is taken is equal to $d(t+1)+\sum_{i=1}^{n_s}a_i(t+1)-g(t+1),$ and this energy must be produced by the reserves. Hence, the stage cost at time $t+1$ equals
\begin{align}
    c_{t+1}(\mathbf{s}(t&), \mathbf{a}(t+1))\nonumber\\
    &=
    c^s\big(g(t+1)-d(t+1)-\sum_{i=1}^{n_s}a_i(t+1)\big).\label{stageCostDefn}
\end{align}

\subsection{Terminal costs}
It follows from the discussion in Section \ref{subsect_transitionKernel} that for any EV $i,$ the term $h_i(T)$ in its final state $s^{EV}_i(T)$ specifies the amount of energy stored in it at the time that it disconnects from the grid. This allows us to capture the costs incurred by the EVs by means of a terminal cost function . Specifically, for any state $\mathbf{s}=\big((e_1(T),h_1(T)),\hdots,(e_{n_s}(T),h_{n_s}(T)\big)\in\mathcal{S},$ we define the terminal cost $v_T(\mathbf{s})$ as 
\begin{align}
    v_T(\mathbf{s})\coloneqq\sum_{i=1}^{n_s}-h_i(T),\label{terminalCostDefn}
\end{align}
which equals the sum of the costs incurred by all EVs in the system.

\subsection{Computing an optimal storage policy}
Consider the Markov Decision Process defined by $\big(\mathcal{S},\{\mathcal{A}_\mathbf{s}\},\{P_t(\mathbf{s}'\vert\mathbf{s},\mathbf{a})\},\{c_{t}\},v_T\big).$ That the MDP admits an optimal policy that is deterministic and Markovian can be established using routine arguments from Markov decision theory, see for example \cite[Section 5.4]{TaylorNotes}. Consequently, the MDP can, in principle, be solved optimally using dynamic programming, which uses the following backward recursion to compute the value function and an optimal policy: 
\begin{align}
    v_t(\mathbf{s})=\min_{\mathbf{a}\in\mathcal{A}_{\mathbf{s}}}\big[c_{t+1}(\mathbf{s},\mathbf{a})+\sum_{\mathbf{s}'\in\mathcal{S}}P_{t+1}(\mathbf{s}'\big\vert\mathbf{s},\mathbf{a})v_{t+1}(\mathbf{s}')\big]\label{DP}
\end{align}
for all $\mathbf{s}\in\mathcal{S}$ and $t=T-1,\hdots,0$. It is easy to verify that $c^g(\mathbf{g})+v_0(\mathbf{s}(0))=\overline{\beta}(\mathbf{g})$, and that the policy obtained by collecting an argument that minimizes the RHS of (\ref{DP}) at each state and each time is an optimal policy $\boldsymbol{\pi}^\dagger(\mathbf{g})$.


To summarize, the stochastic program (\ref{ISO_OneShot}) when the production function $c^s$ is additively separable reduces to a finite-dimensional optimization program which has embedded in it a dynamic program. We have used this reduction in our simulations to compute an optimal energy dispatch, storage policy and cost savings. 

\begin{figure}
    \centering
    \includegraphics[width=\columnwidth]{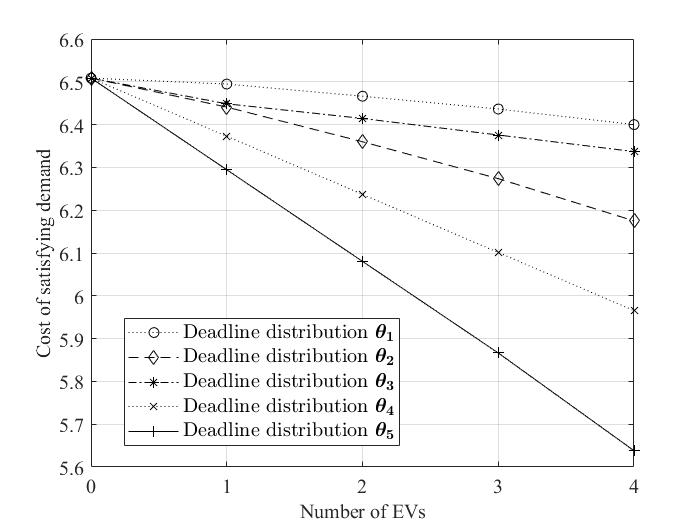}
    \caption{The total expected cost of satisfying the demand is plotted as a function of varying levels of EV penetration for five different profiles of EV deadline distributions. The solid curve corresponds to the scenario where all EVs can lease their battery to the grid for the entire day, and denotes the minimum possible cost at which the ISO can satisfy the demand.}
    \label{FigSW}
\end{figure}

In our simulations, we have divided a day into a total of five time intervals. The simulation parameters are specified in Table \ref{simulationData}. In order to obtain conservative estimates of the cost reduction, we have assumed all EVs to have a capacity of $10\mathrm{kWh},$ which is an order of magnitude lesser than the battery capacity of many EVs that are commercially available today.
With the view of rendering the dynamic program to have a finite state space and action set, we have quantized the energy levels to which the EVs can be charged. Specifically, we have assumed $\mathcal{H}_i=\{0,10\mathrm{kWh}\}$. We have also quantized the generator's energy output so that at any time, the generator can only produce an integer multiple of $10\;\mathrm{kWh}$. This renders the outer program to be finite. In our simulations, we have solved the outer program using a combination of heuristics and exhaustive search. 

The production functions that are used in the simulations are
\begin{align*}
    c^g(\mathbf{g})=\sum_{t=1}^5c_t^g(t)g(t)
\end{align*}
and
\begin{align*}
    c^s(\mathbf{g}_s)=\sum_{t=1}^5\big[c^s_tg_s(t)\mathds{1}_{\{g_s(t)\geq0\}}+c^s_tg_s^2(t)\mathds{1}_{\{g_s(t)<0\}}\big]
\end{align*}
where the numerical values for $[c_1^g\;\hdots\;c^g_5]$ and $[c^s_1\;\hdots\;c^s_5]$ are specified in Table \ref{simulationData}. The coefficients $c^g_1,\;\hdots\;, c^g_5$ have been drawn at random and sorted, and the range from which they have been drawn has been obtained from the range of locational marginal prices at which electricity was traded in October 2020 in ISO New England's day-ahead market \cite{ISO-NEMarketReports}. The function $c^s$ can be thought of as indicating that it is in some sense easier for the reserves to produce energy in real time than to consume it.

Fig. \ref{FigSW} plots the total cost $q^*(\boldsymbol{\theta})$ at which the ISO satisfies the demand for four different deadline distribution profiles $\boldsymbol{\theta}$ of the EVs. In all cases, the expected cost of satisfying the demand decreases almost linearly with EV penetration in the considered regime. Moreover, the distribution $\theta_D$ corresponds to larger deadlines than the distribution $\theta_C$ in that $F_{\theta_D}(t)\leq F_{\theta_C}(t)$ for all $t\in\{1,\hdots,5\}$. Similarly, we also have $F_{\theta_D}(t)\leq F_{\theta_B}(t)\leq F_{\theta_A}(t)$ for all $t\in\{1,\hdots,5\}$. Hence, in addition to showing the reduction in costs due to increased EV penetration, Fig. \ref{FigSW} also illustrates the value that larger EV deadlines provide in reducing the operating cost. Note that these cost reductions are only lower bounds in that solving the dynamic program and the outer program with finer or even no discretization would result in cost reductions that are higher than those indicated in Fig. \ref{FigSW}. 

{Finally and most importantly, these cost reductions can be attained only if the EVs bid their deadline distributions and deadline realizations truthfully, which in turn is guaranteed only in the presence of a mechanism that renders truth-telling a dominant strategy both in the day-ahead market and in real time. As illustrated in the example in Section \ref{introduction}, in the absence of such a mechanism, strategic behavior of EVs could result in total costs in Fig. \ref{FigSW} that are in excess of $6.5$ for all EV penetration levels and all deadline distributions, implying that the ISO would be better off not utilizing the EVs at all for storage. The magnitude of the cost excess depends on the specifics of each EV's bidding strategy, which in turn could be unpredictable in the absence of a unique dominant strategy.}

\begin{table}[ht]
\centering
\caption{Simulation parameters}\label{simulationData}
\begin{tabular}{| c | c | c |}
\hline
 Variable & Value & Units \\
 \hline
 
 $\mathbf{d}$ & $[36.7387\;   38.5138\;   56.6975\;   73.9188\;   57.6061]$ & kWh \\
 \vspace{1mm}
 $c^g$ & $[12.4198\;   18.8367\;   19.1754\;   31.0088\;   33.3978]$ & \$\big/MWh \\
 \vspace{1mm}
 $c^s$ & $[27.8936\;   28.2861\;   29.3702\;   30.5788\;   34.3765]$ & \$\big/MWh \\
 \vspace{1mm}
 ${\theta}_A$ & $[0.2000\;\;0.2000\;\;0.2000\;\;0.2000\;\;0.2000]$ &  \\
 \vspace{1mm}
 ${\theta}_B$ & $[0.0770\;\;    0.2442\;\;    0.0783\;\;    0.0716\;\;    0.5290]$ &  \\
 \vspace{1mm}
 ${\theta}_C$ & $[0.0378\;\;    0.2430\;\;    0.1449\;\;    0.5683\;\;    0.0059]$ &  \\
 \vspace{1mm}
 ${\theta}_D$ & $[0.0212\;\;    0.0462\;\;    0.1019\;\;    0.2061\;\;    0.6245]$ &  \\
 \vspace{1mm}
 ${\theta}_E$ & $[0\;\;    0\;\;    0\;\;    0\;\;    1]$ &  \\
 \vspace{1mm}
 $\boldsymbol{\theta}_1$ & $[\theta_A\;\;\theta_A\;\;\theta_A\;\;\theta_A]$ &  \\ 
 \vspace{1mm}
 $\boldsymbol{\theta}_2$ & $[\theta_B\;\;\theta_B\;\;\theta_B\;\;\theta_B]$ &  \\  
 \vspace{1mm}
 $\boldsymbol{\theta}_3$ & $[\theta_C\;\;\theta_C\;\;\theta_C\;\;\theta_C]$ &  \\
 \vspace{1mm}
 $\boldsymbol{\theta}_4$ & $[\theta_D\;\;\theta_D\;\;\theta_D\;\;\theta_D]$ &  \\
 \vspace{1mm}
 $\boldsymbol{\theta}_5$ & $[\theta_E\;\;\theta_E\;\;\theta_E\;\;\theta_E]$ &  \\
 \hline
\end{tabular}
\end{table}

\section{Extensions}\label{extensions}

We have assumed the deadline distributions and the demand sequence to remain the same on all days. An immediate extension is to address the scenario where the deadline distributions of the EVs and the demand sequence of the load could potentially be different on different days. Denote by $\{\boldsymbol{\theta}(1),\boldsymbol{\theta}(2),\hdots\}$ the sequence of EV parameters on different days, $\boldsymbol{\theta}(l)\in\Theta^{n_s},$ and by $\{\mathbf{d}_1,\mathbf{d}_2,\hdots\}$ the demand sequence on different days. If the number of distinct elements in $\{\boldsymbol{\theta}(1),\boldsymbol{\theta}(2),\hdots\}$ and $\{\mathbf{d}_1,\mathbf{d}_2,\hdots\}$ are finite, then the results of this paper can be readily extended to this more general setting. Specifically, by requiring the EVs to report their parameters in the day-ahead market on all days, categorizing each day into one of a finite number of ``bins" such that the reported deadline distributions and the demand sequence remain the same on all days belonging to a bin, and instantiating in parallel the mechanism presented in the paper -- one for each bin -- we obtain a mechanism that has the desired incentive and optimality properties. 


We have also assumed that the deadlines of all EVs are independent random variables. Relaxing this assumption and developing an analogous mechanism for the case where the EV deadlines could be correlated is an important generalization since it could more accurately model the usage patterns of EVs in real world. Extending the results to the context of multi-bus power systems is another important generalization. 


\section{Conclusion}\label{conclusion}
We have considered the problem of integrating a fleet of strategic EVs with random deadlines into the grid and utilizing them for energy storage. Without appropriate incentive structures, EV-power grid integration could potentially be counterproductive to the cost- and energy-efficient operation of the grid. This fundamentally arises because of two phenomena operating in tandem -- the randomness of EV usage patterns and the possibility of their strategic behavior. We have shown how this problem can be addressed by means of a carefully-designed energy storage market. Specifically, we have designed a mechanism for energy storage markets that guarantees certain incentive and optimality properties. The mechanism allows the ISO to achieve efficient EV-power grid integration and satisfy the demand at minimum possible cost. 

We have also presented a dynamic programming-based algorithm using which the ISO can compute its day-ahead market decisions in an EV-integrated power system. This algorithm has been used to obtain certain numerical results that demonstrate the cost benefits that EV storage services -- unreliable though they may be -- offer the ISO. A few extensions have also been outlined.

\bibliographystyle{IEEEtran}
\bibliography{references.bib}

\begin{appendices}

\section{Proof of Theorem 1}\label{TheoremProof}
\begin{proof}
Arbitrarily fix $i\in\{1,\hdots,n_s\}$ and $\theta_i\in\Theta.$ We begin the proof with four lemmas. 
\begin{lemma}
There exists $\mathcal{E}_i\subset\{1,\hdots,T\}^\infty$ with $\mathbb{P}_{\theta_i}^\infty(\mathcal{E}_i)=0$ such that for all $\delta_i^\infty\notin\mathcal{E}_i,$
\begin{align}
    \lim_{L\to\infty}\frac{1}{L}\sum_{l=1}^{L}J_p(l)\mathds{1}_{\{E_i(l,\delta_i^l,\theta_i)\}}=0.\label{honesty0penalty}
\end{align}\label{Lemma_honesty}
\end{lemma}
\begin{proof}
See Appendix \ref{Lemma1Proof}.
\end{proof}

The next lemma establishes the monotonicity of the optimal cost function $q^*$ in a certain sense.
\begin{lemma}
Let $\lambda_i,\widetilde{\lambda}_i\in\Theta$ be any two parameters such that $F_{\widetilde{\lambda}_i}(t)\geq F_{{\lambda}_i}(t)$ for all $t\in\{1,\hdots,T\}.$ Then,
\begin{align*}
    q^*(\widetilde{\lambda}_i,\boldsymbol{\phi}_{-i})\geq q^*({\lambda}_i,\boldsymbol{\phi}_{-i})
\end{align*}
for all $\boldsymbol{\phi}_{-i}\in\Theta^{n_s-1}.$\label{monotonic}
\end{lemma}

\begin{proof}
See Appendix \ref{Lemma2Proof}.
\end{proof}

\begin{lemma}
Suppose that $\phi_i$ is such that 
\begin{align}
    \alpha(\theta_i,\phi_i)\coloneqq\sup_{t\in\{1,\hdots,T\}}\{F_{\theta_i}(t)-F_{\phi_i}(t)\}>0.\label{alphaDefn}
\end{align}
Then, 
\begin{enumerate}
    \item for some finite $K\geq0,$ 
\begin{align}
    q^*(\theta_i,\boldsymbol{\phi}_{-i})-q^*(\phi_i,\boldsymbol{\phi}_{-i})\leq K\alpha(\theta_i,\phi_i).\label{lipschitzProp}
\end{align}
for all $\boldsymbol{\phi}_{-i}\in\Theta^{n_s-1}.$
\item There exists $\mathcal{E}_i\subset\{1,\hdots,T\}^\infty$ with $\mathbb{P}_{\theta_i}^\infty(\mathcal{E}_i)=0$ such that for all $\delta_i^\infty\notin\mathcal{E}_i,$  
\begin{align}
    \lim_{L\to\infty}\frac{1}{L}\sum_{l=1}^l\mathds{1}_{\{\widehat{\delta}_i(l)>\delta_i(l)\}}\geq \alpha(\theta_i,\phi_i)\label{deadlinemissedoften}
\end{align}
whenever $\widehat{\delta}_i^\infty$ is such that $\sum_{l=1}^\infty\mathds{1}_{\{E_i(l,\widehat{\delta}_i^l,\phi_i)\}}<\infty.$
\end{enumerate}\label{LemmaLipschitz}
\end{lemma}
\begin{proof}
See Appendix \ref{Lemma3Proof}.
\end{proof}

\begin{lemma}
If $\widehat{\delta}_i^\infty$ such that $\sum_{l=1}^\infty\mathds{1}_{\{E_i(l,\widehat{\delta}_i^l,\phi_i)\}}=\infty,$ then, 
\begin{align}
    \limsup_{L\to\infty}{\frac{1}{L}\sum_{l=1}^LJ_p(l)\mathds{1}_{\{E_i(l,\widehat{\delta}_i^l,\phi_i)\}}}=\infty.\label{infinitePenalty}
\end{align}\label{lemmaInfinitePenalty}
\end{lemma}
\begin{proof}
See Appendix \ref{Lemma4Proof}.
\end{proof}

We are now ready to prove the theorem. We first have 
\begin{align*}
    u^\infty_i(&\phi_i,\widehat{\delta}_i^\infty,\boldsymbol{\phi}_{-i},\widehat{\boldsymbol{\delta}}_{-i}^\infty,\omega^\infty_{\boldsymbol{\pi}^*(\boldsymbol{\phi})})
    \nonumber\\
    &=p_i^{DA}(\phi_i,\boldsymbol{\phi}_{-i})&\nonumber\\
    &\;\;\;\;\;\;+\liminf_{L\to\infty}\frac{1}{L}\sum_{l=1}^{L}\Big[p_i^{S}(l,\boldsymbol{\phi},\widehat{\delta}_i^l,\widehat{\boldsymbol{\delta}}(l),\omega_{\boldsymbol{\pi}^*(\boldsymbol{\phi})}(l))\nonumber\\
    &\;\;\;\;\;\;\;\;\;\;\;\;\;-c_i^{EV}\big(\delta_i(l),\widehat{\delta}_i(l),{\boldsymbol{\pi}}_i^*(\widehat{\boldsymbol{\delta}}(l),\omega_{\boldsymbol{\pi}^*(\boldsymbol{\phi})}(l);\boldsymbol{\phi}\big)\Big].
\end{align*}
Substituting (\ref{EVCost}), (\ref{POE}) and (\ref{piRT}) in the above equality and carrying out some algebra yields
\begin{align}
    u^\infty_i&(\phi_i,\widehat{\delta}_i^\infty,\boldsymbol{\phi}_{-i},\widehat{\boldsymbol{\delta}}_{-i}^\infty,\omega^\infty_{\boldsymbol{\pi}^*(\boldsymbol{\phi})})\nonumber\\
    &=\big[q^*(\boldsymbol{\phi}_{-i})-q^*(\boldsymbol{\phi})\big]\nonumber\\
    &\;\;\;\;-\limsup_{L\to\infty}\frac{1}{L}\sum_{l=1}^{L}\Big[J_p(l)\mathds{1}_{\{E_i(l,\widehat{\delta}_i^l,\phi_i)\}}\nonumber\\
    &\;\;\;\;\;\;\;\;\;\;\;\;\;\;\;\;\;\;\;\;\;\;\;\;\;\;\;\;\;\;\;\;\;\;+J_m\mathds{1}_{\{\widehat{\delta}_i(l)>\delta_i(l)\}}\Big]\nonumber\\
    &\;\;\;\;-\limsup_{L\to\infty}\frac{1}{L}\sum_{l=1}^{L}\Big[{\pi}_i^*(\widehat{\delta}_i(l),\widehat{\boldsymbol{\delta}}(l),\omega_{\boldsymbol{\pi}^*(\boldsymbol{\phi})}(l);\boldsymbol{\phi})\nonumber\\
    &\;\;\;\;\;\;\;\;\;-\pi_i^*(\widehat{\delta}_i(l),\widehat{\boldsymbol{\delta}}(l),\omega_{\boldsymbol{\pi}^*(\boldsymbol{\phi})}(l);\boldsymbol{\phi})\mathds{1}_{\{\widehat{\delta}_i(l)\leq\delta_i(l)\}}\Big].\label{utilityexpression}
\end{align}
Using the above expression, applying Lemma \ref{Lemma_honesty}, and carrying out some algebra implies the existence of $\mathcal{E}_i\subset\{1,\hdots,T\}^\infty$ with $\mathbb{P}_{\theta_i}^\infty(\mathcal{E}_i)=0$ such that for all $\delta_i^\infty\notin\mathcal{E}_i,$
\begin{align}
    u^\infty_i&(\theta_i,{\delta}_i^\infty,\boldsymbol{\phi}_{-i},\widehat{\boldsymbol{\delta}}_{-i}^\infty,\omega^\infty_{\boldsymbol{\pi}^*([\theta_i,\boldsymbol{\phi}_{-i}])})\nonumber\\
    &-u^\infty_i(\phi_i,\widehat{\delta}_i^\infty,\boldsymbol{\phi}_{-i},\widehat{\boldsymbol{\delta}}_{-i}^\infty,\omega^\infty_{\boldsymbol{\pi}^*(\boldsymbol{\phi})})\nonumber\\
    &\;\;\;\;=\big[q^*(\phi_i,\boldsymbol{\phi}_{-i})-q^*(\theta_i,\boldsymbol{\phi}_{-i})\big]\nonumber\\
    &\;\;\;\;\;\;\;\;+\limsup_{L\to\infty}\frac{1}{L}\sum_{l=1}^{L}\Big[J_p(l)\mathds{1}_{\{E_i(l,\widehat{\delta}_i^l,\phi_i)\}}\Big]\nonumber\\
    &\;\;\;\;\;\;\;\;\;\;\;\;+\limsup_{L\to\infty}\frac{1}{L}\sum_{l=1}^{L}\Big[{\pi}_i^*(\widehat{\delta}_i(l),\widehat{\boldsymbol{\delta}}(l),\omega_{\boldsymbol{\pi}^*(\boldsymbol{\phi})}(l);\boldsymbol{\phi})\nonumber\\
    &\;\;\;\;\;\;\;\;\;\;\;\;\;\;\;\;-\pi_i^*(\widehat{\delta}_i(l),\widehat{\boldsymbol{\delta}}(l),\omega_{\boldsymbol{\pi}^*(\boldsymbol{\phi})}(l);\boldsymbol{\phi})\mathds{1}_{\{\widehat{\delta}_i(l)\leq\delta_i(l)\}}\Big]\nonumber\\
    &\;\;\;\;\;\;\;\;\;\;\;\;\;\;\;\;\;\;\;\;\;\;\;\;+\limsup_{L\to\infty}\frac{1}{L}\sum_{l=1}^{L}J_m\mathds{1}_{\{\widehat{\delta}_i(l)>\delta_i(l)\}}.\label{utilityDiff1}
\end{align}

We first show that the above random variable is nonnegative, thereby establishing (\ref{DSIC}). We do this by considering three cases and showing that (\ref{utilityDiff1}) is nonnegative in all the three cases.

We first consider the case when $\widehat{\delta}_i^\infty$ is such that $\sum_{l=1}^\infty\mathds{1}_{\{E_i(l,\widehat{\delta}_i^l,\phi_i)\}}=\infty.$ Note that $\big\vert q^*(\phi_i,\boldsymbol{\phi}_{-i})-q^*(\theta_i,\boldsymbol{\phi}_{-i})\big\vert<\infty$ owing to (\ref{costFinite}), and that all other terms in the RHS of (\ref{utilityDiff1}) are nonnegative. Consequently, substituting (\ref{infinitePenalty}) in (\ref{utilityDiff1}) implies the nonnegativity of the (\ref{utilityDiff1}).

We next consider the case when $\phi_i$ is such that $F_{\theta_i}(t)\leq F_{\phi_i}(t)$ for all $t\in\{1,\hdots,T\}.$ It then follows from Lemma \ref{monotonic} that $q^*(\phi_i,\boldsymbol{\phi}_{-i})\geq q^*(\theta_i,\boldsymbol{\phi}_{-i}),$ and so the first term in the RHS of (\ref{utilityDiff1}) is nonnegative. Since every other term in the RHS of (\ref{utilityDiff1}) is nonnegative, (\ref{utilityDiff1}) is nonnegative.

We are finally left with the case when $\widehat{\delta}_i^\infty$ is such that $\sum_{l=1}^\infty\mathds{1}_{\{E_i(l,\widehat{\delta}_i^l,\phi_i)\}}<\infty$ and $\phi_i$ is such that $F_{\theta_i}(t)>F_{\phi_i}(t)$ for some $t\in\{1,\hdots,T\}.$ Lemma \ref{LemmaLipschitz} applies, and so combining (\ref{lipschitzProp}) and (\ref{deadlinemissedoften}) with (\ref{utilityDiff1}) implies that $\mathbb{P}_{\theta_i}^\infty-$ almost surely,
\begin{align*}
    u^\infty_i(\theta_i,{\delta}_i^\infty,\boldsymbol{\phi}_{-i},\widehat{\boldsymbol{\delta}}_{-i}^\infty,\omega^\infty_{\boldsymbol{\pi}^*([\theta_i,\boldsymbol{\phi}_{-i}])}&)\nonumber\\
    -u^\infty_i(\phi_i,\widehat{\delta}_i^\infty,\boldsymbol{\phi}_{-i},\widehat{\boldsymbol{\delta}}_{-i}^\infty,&\omega^\infty_{\boldsymbol{\pi}^*(\boldsymbol{\phi})})\nonumber\\
    \geq (J_m&-K)\alpha(\theta_i,\phi_i).
\end{align*}
It follows that for $J_m\geq K,$ the LHS is nonnegative, thereby completing the proof of (\ref{DSIC}). 

We next prove the second statement of the theorem. Suppose that (\ref{equalUtilities}) holds. 
Using (\ref{utilityexpression}) to expand both sides of (\ref{equalUtilities}), invoking (\ref{honesty0penalty}), and simplifying the result implies that for every $\delta_i^\infty\notin\mathcal{E}_i$, $\boldsymbol{\phi}_{-i},$ $\widehat{\boldsymbol{\delta}}^\infty,$ $\omega^\infty_{\boldsymbol{\pi}^*(\boldsymbol{\phi})},\omega^\infty_{\boldsymbol{\pi}^*([\theta_i,\boldsymbol{\phi}_{-i}])}$, there exists $\widehat{\delta}_i^\infty$ such that, 
\begin{align}
    q^*(&\theta_i,\boldsymbol{\phi}_{-i})-q^*(\phi_i,\boldsymbol{\phi}_{-i})\nonumber\\
    &=\limsup_{L\to\infty}\frac{1}{L}\sum_{l=1}^LJ_p(l)\mathds{1}_{E_i(l,\widehat{\delta}_i^l,\phi_i)}\nonumber\\
    &\;\;\;\;\;\;+\limsup_{L\to\infty}\frac{1}{L}\sum_{l=1}^L\Big[{\pi}_i^*(\widehat{\delta}_i(l),\widehat{\boldsymbol{\delta}}(l),\omega_{\boldsymbol{\pi}^*(\boldsymbol{\phi})}(l);\boldsymbol{\phi})\nonumber\\
    &\;\;\;\;\;\;\;\;\;\;\;\;\;-\pi_i^*(\widehat{\delta}_i(l),\widehat{\boldsymbol{\delta}}(l),\omega_{\boldsymbol{\pi}^*(\boldsymbol{\phi})}(l);\boldsymbol{\phi})\mathds{1}_{\{\widehat{\delta}_i(l)\leq\delta_i(l)\}}\Big]\nonumber\\
    &\;\;\;\;\;\;\;\;\;\;\;\;\;\;\;\;\;\;\;\;\;\;+J_m\limsup_{L\to\infty}\frac{1}{L}\sum_{l=1}^L\mathds{1}_{\{\widehat{\delta}_i(l)>\delta_i(l)\}}.\label{equalUtilityConsequence}
\end{align}
Since the RHS of the above equality is always nonnegative, we have that $q^*(\theta_i,\boldsymbol{\phi}_{-i})-q^*(\phi_i,\boldsymbol{\phi}_{-i})\geq0$ for every $\boldsymbol{\phi}_{-i}.$ We show next that this inequality must hold with equality for all $\boldsymbol{\phi}_{-i}$. 

Suppose for contradiction that $q^*(\theta_i,\boldsymbol{\phi}_{-i})-q^*(\phi_i,\boldsymbol{\phi}_{-i})>0$ for some $\boldsymbol{\phi}_{-i}.$ Lemma \ref{monotonic} implies that $\sup_{t\in\{1,\hdots,T\}}\{F_{\theta_i}(t)-F_{\phi_i}(t)\}=\alpha(\theta_i,\phi_i)>0$, and so Lemma \ref{LemmaLipschitz} applies. Hence,
\begin{align}
    q^*(\theta_i,\boldsymbol{\phi}_{-i})-q^*(\phi_i,\boldsymbol{\phi}_{-i})\leq K\alpha(\theta_i,\phi_i).\label{p3u1}
\end{align}
Since the LHS of (\ref{equalUtilityConsequence}) is finite following (\ref{costFinite}), the RHS must also be finite. This implies in particular that the first term of the RHS is finite, and so, it follows from Lemma \ref{lemmaInfinitePenalty} that $\sum_{l=1}^\infty\mathds{1}_{\{E_i(l,\widehat{\delta}_i^l,\phi_i)\}}<\infty.$ Consequently, (\ref{deadlinemissedoften}) holds, and substituting it in (\ref{equalUtilityConsequence}) implies that $\mathbb{P}_{\theta_i}^\infty-$ almost surely,
\begin{align}
    q^*(\theta_i,\boldsymbol{\phi}_{-i})-q^*(\phi_i,\boldsymbol{\phi}_{-i})\geq J_m\alpha(\theta_i,\phi_i).\label{p3u2}
\end{align}
Combining (\ref{p3u1}) and (\ref{p3u2}) yields a contradiction for $J_m>K.$ Hence, if (\ref{equalUtilities}) holds, then, 
\begin{align}
    q^*(\theta_i,\boldsymbol{\phi}_{-i})=q^*(\phi_i,\boldsymbol{\phi}_{-i}).\label{DACostsEqual}
\end{align}
for all $\boldsymbol{\phi}_{-i}.$ 

Now, since (\ref{uniqueDScondition}) holds, combining it with (\ref{DACostsEqual}) immediately establishes (\ref{thetaTruthful}), i.e., that $\phi_i=\theta_i.$ Consequently, the RHS of (\ref{equalUtilityConsequence}) is zero, and since every term in the RHS of (\ref{equalUtilityConsequence}) is nonnegative, it follows that every term is zero. Hence, 
\begin{align}
    \lim_{L\to\infty}\frac{1}{L}\sum_{l=1}^LJ_p(l)\mathds{1}_{E_i(l,\widehat{\delta}_i^l,\theta_i)}=0\label{p2p2u1}
\end{align}
and 
\begin{align}
    \lim_{L\to\infty}\frac{1}{L}\sum_{l=1}^L\mathds{1}_{\{\widehat{\delta}_i(l)>\delta_i(l)\}}=0\label{p2p2u2}
\end{align}
$\mathbb{P}_{\theta_i}^\infty-$ almost surely.

From (\ref{p2p2u1}), we get $\lim_{L\to\infty}\frac{1}{L}\sum_{l=1}^L\mathds{1}_{E_i(l,\widehat{\delta}_i^l,\theta_i)}=0,$ which implies using (\ref{Edefn}) that for every $t\in\{1,\hdots,T\},$  
\begin{align}
    \lim_{L\to\infty}\frac{1}{L}\sum_{l=1}^L\mathds{1}_{\{\vert{f}_{i,t}(l,\widehat{\delta}_i^l,\theta_i)\vert\geq r(l)\}}=0.\label{fhatRarelyExceeds}
\end{align}
Since $\vert{f}_{i,t}(l,\widehat{\delta}_i^l,\theta_i)\vert\leq1,$ we have that for all $t\in\{1,\hdots,T\}$ and all $L\in\mathbb{N},$ 
\begin{align*}
    \sum_{l=1}^L{\vert{f}_{i,t}(l,\widehat{\delta}_i^l,\theta_i)\vert}\leq\sum_{l=1}^L&\mathds{1}_{\{\vert{f}_{i,t}(l,\widehat{\delta}_i^l,\theta_i)\vert\geq r(l)\}}\nonumber\\
    &+\sum_{l=1}^Lr(l)\mathds{1}_{\{\vert{f}_{i,t}(l,\widehat{\delta}_i^l,\theta_i)\vert< r(l)\}}.
\end{align*}
Dividing this inequality by $L$ and taking the limit as $L\to\infty$ implies that for all $t\in\{1,\hdots,T\},$ 
\begin{align*}
    \limsup_{L\to\infty}\frac{1}{L}\sum_{l=1}^L&{\vert{f}_{i,t}(l,\widehat{\delta}_i^l,\theta_i)\vert}\nonumber\\
    \leq\limsup_{L\to\infty}&\frac{1}{L}\sum_{l=1}^L\mathds{1}_{\{\vert{f}_{i,t}(l,\widehat{\delta}_i^l,\theta_i)\vert\geq r(l)\}}\nonumber\\
    &+\limsup_{L\to\infty}\frac{1}{L}\sum_{l=1}^Lr(l)\mathds{1}_{\{\vert{f}_{i,t}(l,\widehat{\delta}_i^l,\theta_i)\vert< r(l)\}}.
\end{align*}
We have using (\ref{rgoestozero}) that the last term of the RHS of the above inequality equals zero, and using (\ref{fhatRarelyExceeds}) that its first term is zero. Hence, for all $t\in\{1,\hdots,T\},$ we have $\lim_{L\to\infty}\frac{1}{L}\sum_{l=1}^L{\vert{f}_{i,t}(l,\widehat{\delta}_i^l,\theta_i)\vert}=0.$ This implies using (\ref{fdefn}) that for all $t\in\{1,\hdots,T\},$ $\lim_{L\to\infty}\frac{1}{L}\sum_{l=1}^L{\bigg[\frac{1}{l}\sum_{l'=1}^{l}\mathds{1}_{\{\widehat{\delta}_i(l')=t\}}\bigg]}=\mathbb{P}_{\theta_i}(t).$
Multiplying this equality by $t$, summing both sides over $t$, and simplifying yields
\begin{align}
    \lim_{L\to\infty}\frac{1}{L}\sum_{l=1}^L\bigg[\frac{1}{l}{\sum_{l'=1}^{l}\widehat{\delta}_i(l')}\bigg]=\mu(\theta_i),\label{cesaroConvergence}
\end{align}
where $\mu(\theta_i)\coloneqq\sum_{t=1}^Tt\mathbb{P}_{\theta_i}(t)$ is the expected value of the distribution corresponding to the parameter $\theta_i.$ 
Next, we use this to show that $\lim_{L\to\infty}\frac{1}{L}\sum_{l=1}^L\mathds{1}_{\{\widehat{\delta}_i(l)<\delta_i(l)\}}=0,$ which when combined with (\ref{p2p2u2}) would establish (\ref{deltaTruthful}).

Suppose for contradiction that for some $\epsilon>0,$
\begin{align}
    \limsup_{L\to\infty}\frac{1}{L}\sum_{l=1}^L\mathds{1}_{\{\widehat{\delta}_i(l)<\delta_i(l)\}}=\epsilon.\label{AssumeContra}
\end{align}
Note that
\begin{align}
    &\limsup_{L\to\infty}\bigg[\frac{1}{L}\sum_{l=1}^{L}\widehat{\delta}_i(l)\bigg]\nonumber\\
    &\;\;\:\;\;\;=\limsup_{L\to\infty}\bigg[\frac{1}{L}{\sum_{l=1}^{L}\widehat{\delta}_i(l)}\mathds{1}_{\{\widehat{\delta}_i(l)=\delta_i(l)\}}\bigg]\nonumber\\
    &\;\;\;\;\;\;\;\;\;\;\;\;\;\;+\limsup_{L\to\infty}\bigg[\frac{1}{L}{\sum_{l=1}^{L}\widehat{\delta}_i(l)}\mathds{1}_{\{\widehat{\delta}_i(l)<\delta_i(l)\}}\bigg]\nonumber\\
    &\;\;\;\;\;\;\;\;\;\;\;\;\;\;\;\;\;\;\;\;\;\;\;+\limsup_{L\to\infty}\bigg[\frac{1}{L}{\sum_{l=1}^{L}\widehat{\delta}_i(l)}\mathds{1}_{\{\widehat{\delta}_i(l)>\delta_i(l)\}}\bigg].\label{p2p2u4}
\end{align}
Now, 
\begin{align*}
    \limsup_{L\to\infty}\bigg[\frac{1}{L}\sum_{l=1}^{L}&\widehat{\delta}_i(l)\mathds{1}_{\{\widehat{\delta}_i(l)<\delta_i(l)\}}\bigg]\nonumber\\
    <\limsup_{L\to\infty}\bigg[\frac{1}{L}&{\sum_{l=1}^{L}\widehat{\delta}_i(l)}\mathds{1}_{\{\widehat{\delta}_i(l)<\delta_i(l)\}}\bigg]+\epsilon\nonumber\\
    =\limsup_{L\to\infty}&\bigg[\frac{1}{L}{\sum_{l=1}^{L}\big(\widehat{\delta}_i(l)+1\big)}\mathds{1}_{\{\widehat{\delta}_i(l)<\delta_i(l)\}}\bigg]\nonumber\\
    &\leq\limsup_{L\to\infty}\bigg[\frac{1}{L}{\sum_{l=1}^{L}{\delta}_i(l)}\mathds{1}_{\{\widehat{\delta}_i(l)<\delta_i(l)\}}\bigg],
\end{align*}
where the equality follows from (\ref{AssumeContra}). 
It follows from (\ref{p2p2u2}) that the last term in the RHS of (\ref{p2p2u4}) is $\mathbb{P}_{\theta_i}^\infty-$ almost surely zero, and so substituting this and the above inequality in (\ref{p2p2u4}) implies that $\mathbb{P}_{\theta_i}^\infty-$ almost surely,
\begin{align*}
    \limsup_{L\to\infty}\bigg[\frac{1}{L}{\sum_{l=1}^{L}\widehat{\delta}_i(l)}\bigg]<\limsup_{L\to\infty}\bigg[\frac{1}{L}{\sum_{l=1}^{L}{\delta}_i(l)}\mathds{1}_{\{\widehat{\delta}_i(l)\leq\delta_i(l)\}}\bigg].
\end{align*}
It follows from (\ref{p2p2u2}) and SLLN that the RHS of the above inequality $\mathbb{P}_{\theta_i}^\infty-$ almost surely equals $\mu(\theta_i)$, and so, $\limsup_{L\to\infty}\frac{1}{L}{\sum_{l=1}^{L}\widehat{\delta}_i(l)}<\mu(\theta_i).$
Combining this inequality with (\ref{cesaroConvergence}) yields a contradiction. Hence, $\lim_{L\to\infty}\frac{1}{L}\sum_{l=1}^L\mathds{1}_{\{\widehat{\delta}_i(l)<\delta_i(l)\}}=0,$ and combining this with (\ref{p2p2u2}) establishes (\ref{deltaTruthful}).


We now prove the third part of the theorem. Using (\ref{utilityexpression}) to compute $u^\infty_i(\theta_i,{\delta}_i^\infty,\boldsymbol{\phi}_{-i},\widehat{\boldsymbol{\delta}}_{-i}^\infty,\omega^\infty_{\boldsymbol{\pi}^*(\theta_i,\boldsymbol{\phi}_{-i})})$ yields
\begin{align*}
    u^\infty_i(\theta_i,{\delta}_i^\infty,&\boldsymbol{\phi}_{-i},\widehat{\boldsymbol{\delta}}_{-i}^\infty,\omega^\infty_{\boldsymbol{\pi}^*(\theta_i,\boldsymbol{\phi}_{-i})})\nonumber\\
    &=\big[q^*(\boldsymbol{\phi}_{-i})
    -q^*(\theta_i,\boldsymbol{\phi}_{-i})\big]
    \nonumber\\
    &\;\;\;\;\;\;\;\;-\limsup_{L\to\infty}\frac{1}{L}\sum_{l=1}^{L}\Big[J_p(l)\mathds{1}_{\{E_i(l,{\delta}_i^l,\theta_i)\}}\Big].
\end{align*}
Substituting (\ref{honesty0penalty}) in the above equality and noting that $[q^*(\boldsymbol{\phi}_{-i})-q^*(\theta_i,\boldsymbol{\phi}_{-i})]\geq0$ establishes (\ref{IR}). 

Finally, (\ref{efficiencyOfMarket}) follows by substituting (\ref{thetaTruthful}) in the LHS of (\ref{efficiencyOfMarket}) and using (\ref{deltaTruthful}) and SLLN to simplify the result.
\end{proof}

\section{Proof of Lemma 1}\label{Lemma1Proof}
\begin{proof}
Denote by $\sigma^2_{t}=\mathbb{P}_{\theta_i}(t)\big(1-\mathbb{P}_{\theta_i}(t)\big)$ the variance of the random variable $\mathds{1}_{\{\delta_i(1)=t\}}-\mathbb{P}_{\theta_i}(t).$ Since $\delta_i^\infty$ is drawn IID from the distribution $\mathbb{P}_{\theta_i},$ it follows from the Central Limit Theorem (CLT) that for any $z\in\mathbb{R},$
\begin{align}
    \lim_{l\to\infty}\sup_{z\in\mathbb{R}}\Big\vert\mathbb{P}\{\sqrt{l}{f}_{i,t}(l,\delta_i^l,\theta_i)\leq z\}-Q\big(\frac{z}{\sigma_{t}}\big)\Big\vert=0\label{CLT1}
\end{align}
for all $t\in\{1,\hdots,T\},$ where $Q$ denotes the cumulative distribution function of the standard normal distribution. Define 
\begin{align}
    \widetilde{r}(l)\coloneqq r(l)\sqrt{l}.\label{rtildedefn}
\end{align}
It follows from (\ref{CLT1}) that 
\begin{align*}
    \lim_{l\to\infty}\sup_{q\in\mathbb{N}}\Big\vert\mathbb{P}\{\sqrt{l}{f}_{i,t}(l,\delta_i^l,\theta_i)\leq -\widetilde{r}(q)\}-Q\big(\frac{-\widetilde{r}(q)}{\sigma_{t}}\big)\Big\vert=0.
\end{align*}
Hence,
\begin{align*}
    \lim_{l\to\infty}\Big\vert\mathbb{P}\{\sqrt{l}{f}_{i,t}(l,\delta_i^l,\theta_i)\leq -\widetilde{r}(l)\}-Q\big(\frac{-\widetilde{r}(l)}{\sigma_{t}}\big)\Big\vert=0,
\end{align*}
which using (\ref{rtildedefn}) becomes
\begin{align*}
    \lim_{l\to\infty}\Big\vert\mathbb{P}\{{f}_{i,t}(l,\delta_i^l,\theta_i)\leq -{r}(l)\}-Q\big(\frac{-\widetilde{r}(l)}{\sigma_{t}}\big)\Big\vert=0.
\end{align*}
The above equality implies that for every $\epsilon_0>0,$ there exists $L_0\in\mathbb{N}$ such that for all $l\geq L_0,$
\begin{align*}
    -\epsilon_0\leq\mathbb{P}\{{f}_{i,t}(l,\delta_i^l,\theta_i)\leq {-r(l)}\}-Q\big(\frac{-\widetilde{r}(l)}{\sigma_t}\big)\leq\epsilon_0.
\end{align*}
Using the bound $Q(z)\leq e^{-{z^2}/{2}}$ for $z\leq 0,$ the above inequality implies that for all $l\geq L_0,$ 
\begin{align*}
    \mathbb{P}\{{f}_{i,t}(l,\delta_i^l,\theta_i)\leq {-r(l)}\}\leq e^{-{\widetilde{r}^2(l)}/{2\sigma_t^2}}+\epsilon_0.
\end{align*}
It follows from (\ref{romega}) and (\ref{rtildedefn}) that $\widetilde{r}(l)\geq\sqrt{\ln{l^{\gamma}}}$ for all $l\geq L_r.$ Combining this with the above inequality and simplifying yields that for all $l\geq \max{\{L_r,L_0\}},$
\begin{align*}
    \mathbb{P}\{{f}_{i,t}(l,\delta_i^l,\theta_i)\leq {-r(l)}\}\leq 
    \frac{1}{l^{(\frac{\gamma}{2\sigma_t^2})}}+\epsilon_0,
\end{align*}
and so, letting $\epsilon_0\downarrow0$ and noting that $2\sigma_t^2\leq\frac{1}{2}$ implies
\begin{align*}
    \mathbb{P}\{{f}_{i,t}(l,\delta_i^l,\theta_i)\}\leq {-r(l)}\}=O\Big(\frac{1}{l^{2\gamma}}\Big).
\end{align*}
Since $\gamma>\frac{1}{2},$ the above equality implies
\begin{align}
    \sum_{l=1}^\infty\mathbb{P}\{{f}_{i,t}(l,\delta_i^l,\theta_i)\leq {-r(l)}\}<\infty.\label{l1u1}
\end{align}
It can be shown by following the same sequence of arguments that 
\begin{align}
    \sum_{l=1}^\infty\mathbb{P}\{{f}_{i,t}(l,\delta_i^l,\theta_i)\geq {r(l)}\}<\infty.\label{l1u2}
\end{align}
Combining (\ref{l1u1}) and (\ref{l1u2}) implies
\begin{align*}
    \sum_{l=1}^\infty\mathbb{P}\{\big\vert{f}_{i,t}(l,\delta_i^l,\theta_i)\big\vert\geq {r(l)}\}<\infty.
\end{align*}
Since the above inequality holds for arbitrary $t\in\{1,\hdots,T\},$ it follows that $\sum_{l=1}^\infty\mathbb{P}\{E_i(l,\delta_i^l,\theta_i)\}<\infty.$ Invoking the Borel-Cantelli lemma yields
\begin{align*}
    \sum_{l=1}^\infty\mathds{1}_{\{E_i(l,\delta_i^l,\theta_i)\}}<\infty
\end{align*}
almost surely, and (\ref{honesty0penalty}) follows.
\end{proof}

\section{Proof of Lemma 2}\label{Lemma2Proof}
\begin{proof}
Since $F_{\widetilde{\lambda}_i}(t)\geq F_{\lambda_i}(t)$ for every $t\in\{1,\hdots,T\},$ there exists a probability space $(\Omega_{\kappa},\mathcal{F}_{\kappa},\mathbb{P}_{\kappa})$ and a function $f_S:\{1,\hdots,T\}\times\Omega_{\kappa}\to\{1,\hdots,T\}$ such that 
\begin{align}
    f_S(t,\zeta)\leq t\label{deadlineRespected}
\end{align}
for all $t\in\{1,\hdots,T\}$ and all $\zeta\in\Omega_{\kappa}$, and 
\begin{align}
    f_S(t,\zeta)\sim\mathbb{P}_{\widetilde{\lambda}_i}
\end{align}
when $t\sim\mathbb{P}_{\lambda_i}$ and $\zeta\sim\mathbb{P}_{\kappa}.$


Now, fix $\boldsymbol{\phi}_{-i}$ arbitrarily and define a storage policy $\boldsymbol{\nu}$ as follows. For each $j\in\{1,\hdots,n_s\},$ define
\begin{align}
    \nu_j(t,\boldsymbol{\delta},(\zeta,\omega))=
    \pi_j^*\big(t,[f_S(\delta_i,\zeta),\boldsymbol{\delta}_{-i}],\omega\;;[\widetilde{\lambda}_i,\boldsymbol{\phi}_{-i}]\big),\label{nuDefn}
\end{align}
with $(\zeta,\omega)\sim\mathbb{P}_{\kappa}\times\mathbb{P}_{\boldsymbol{\pi}^*([\widetilde{\lambda}_i,\boldsymbol{\phi}_{-i}])}.$ In words, the policy $\boldsymbol{\nu}$ is the result of implementing the policy $\boldsymbol{\pi}^*([\widetilde{\lambda}_i,\boldsymbol{\phi}_{-i}])$ under the pretense that the deadline of EV $i$ is $f_S(\delta_i,\zeta)$ when it is in fact $\delta_i.$ It is straightforward to verify using (\ref{deadlineRespected}) that if the policy $\boldsymbol{\pi}^*([\widetilde{\lambda}_i,\boldsymbol{\phi}_{-i}])$ is implementable, then so is the policy $\boldsymbol{\nu}$. For each $j\in\{1,\hdots,n_s\}$, define 
\begin{align}
    \overline{\nu}_j(t,\boldsymbol{\delta},\zeta)\coloneqq\mathbb{E}_{\omega\sim \mathbb{P}_{\boldsymbol{\pi}^*([\widetilde{\lambda}_i,\boldsymbol{\phi}_{-i}])}}\big[\nu_j(t,\boldsymbol{\delta},(\zeta,\omega))\big]\label{nuBarDefn}
\end{align}
and
\begin{align*}
    \widehat{\nu}_j(t,\boldsymbol{\delta})\coloneqq\mathbb{E}_{(\zeta,\omega)\sim\mathbb{P}_{\kappa}\times\mathbb{P}_{\boldsymbol{\pi}^*([\widetilde{\lambda}_i,\boldsymbol{\phi}_{-i}])}}\big[\nu_j(t,\boldsymbol{\delta},(\zeta,\omega))\big]
\end{align*}
so that 
\begin{align*}
    \widehat{\nu}_j(t,\boldsymbol{\delta})=\mathbb{E}_{\zeta\sim\mathbb{P}_{\kappa}}\big[\overline{\nu}_j(t,\boldsymbol{\delta},\zeta)\big]
\end{align*}
and 
\begin{align}
    \mathbb{E}_{(\boldsymbol{\delta},\zeta)\sim \mathbb{P}_{([\lambda_i,\boldsymbol{\phi}_{-i}])}\times\mathbb{P}_{\kappa}}\big[\overline{\nu}_j(t,\boldsymbol{\delta},\zeta)\big]=\mathbb{E}_{\boldsymbol{\delta}\sim\mathbb{P}_{(\lambda_i,\boldsymbol{\phi}_{-i})}}\big[\widehat{\nu}_j(t,\boldsymbol{\delta})\big].\label{l2u1}
\end{align}
Substituting (\ref{nuDefn}) and (\ref{piHatDefn}) in (\ref{nuBarDefn}), we obtain
\begin{align*}
    \overline{\nu}_j(t,\boldsymbol{\delta},\zeta)=\widehat{\pi}^*_j\big(t,[f_S(\delta_i,\zeta),\boldsymbol{\delta}_{-i}];[\widetilde{\lambda}_i,\boldsymbol{\phi}_{-i}]\big),
\end{align*}
and so 
\begin{align*}
    \mathbb{E}_{(\boldsymbol{\delta},\zeta)\sim \mathbb{P}_{(\lambda_i,\boldsymbol{\phi}_{-i})}\times\mathbb{P}_{\kappa}}\big[\overline{\nu}_j(t,&\boldsymbol{\delta},\zeta)\big]\nonumber\\
    =\mathbb{E}_{(\boldsymbol{\delta},\zeta)\sim \mathbb{P}_{(\lambda_i,\boldsymbol{\phi}_{-i})}\times\mathbb{P}_{\kappa}}&\big[\widehat{\pi}^*_j\big(t,[f_S(\delta_i,\zeta),\boldsymbol{\delta}_{-i}];[\widetilde{\lambda}_i,\boldsymbol{\phi}_{-i}]\big)\big]\nonumber\\
    =&\mathbb{E}_{\boldsymbol{\delta}'\sim \mathbb{P}_{(\widetilde{\lambda}_i,\boldsymbol{\phi}_{-i})}}\big[\widehat{\pi}^*_j(t,\boldsymbol{\delta}';(\widetilde{\lambda}_i,\boldsymbol{\phi}_{-i}))\big],
\end{align*}
where the last equality follows from the fact that $f_S(\delta_i,\zeta)\sim\mathbb{P}_{\widetilde{\lambda}_i}$ and so $[f_S(\delta_i,\zeta),\boldsymbol{\delta}_{-i}]\sim\mathbb{P}_{([\widetilde{\lambda}_i,\boldsymbol{\phi}_{-i}])}$. Substituting this in (\ref{l2u1}) yields
\begin{align}
    \mathbb{E}_{\boldsymbol{\delta}\sim\mathbb{P}_{(\lambda_i,\boldsymbol{\phi}_{-i})}}\big[\widehat{\nu}_j(t,\boldsymbol{\delta})\big]=\mathbb{E}_{\boldsymbol{\delta}'\sim \mathbb{P}_{(\widetilde{\lambda}_i,\boldsymbol{\phi}_{-i})}}\big[\widehat{\pi}^*_j\big(t,\boldsymbol{\delta}';[\widetilde{\lambda}_i,\boldsymbol{\phi}_{-i}]\big)\big].\label{l2u2}
\end{align}

Next, we have for every $t\in\{1,\hdots,T\}$  and every $\boldsymbol{\delta}\in\{1,\hdots,T\}^{n_s},$
\begin{align*}
    g_s(t,\boldsymbol{\delta};\mathbf{g}^*([\widetilde{\lambda}_i,\boldsymbol{\phi}_{-i}]),\boldsymbol{\nu},(\zeta,\omega))\nonumber\\
    =d(t)-g^*(t;[\widetilde{\lambda}_i,\boldsymbol{\phi}_{-i}])&\nonumber\\
    -\sum_{j=1}^{n_s}\bigg[\nu_j\big(t,\boldsymbol{\delta},(\zeta,\omega)\big)&-\nu_j\big(t-1,\boldsymbol{\delta},(\zeta,\omega)\big)\bigg]\nonumber\\
    =d(t)-g^*(t;[\widetilde{\lambda}_i,\boldsymbol{\phi}_{-i}])&\nonumber\\
    -\sum_{j=1}^{n_s}\bigg[\pi^*_j(t,[f_S(\delta_i,\zeta),&\boldsymbol{\delta}_{-i}],\omega;[\widetilde{\lambda}_i,\boldsymbol{\phi}_{-i}])\nonumber\\
    -\pi^*_j(t-1,&[f_S(\delta_i,\zeta),\boldsymbol{\delta}_{-i}],\omega;[\widetilde{\lambda}_i,\boldsymbol{\phi}_{-i}])\bigg]\nonumber\\
    =g_s(t,[f_S(\delta_i,\zeta),\boldsymbol{\delta}_{-i}];\mathbf{g}^*(&[\widetilde{\lambda}_i,\boldsymbol{\phi}_{-i}]),\boldsymbol{\pi}^*([\widetilde{\lambda}_i,\boldsymbol{\phi}_{-i}]),\omega),
\end{align*}
where the first and the last equalities follow from (\ref{gstDefn}), and the second equality follows from (\ref{nuDefn}). Hence, 
\begin{align*}
    c^s(\mathbf{g}_s\big(&\boldsymbol{\delta}\;;\;\mathbf{g}^*([\widetilde{\lambda}_i,\boldsymbol{\phi}_{-i}]),\boldsymbol{\nu},(\zeta,\omega)\big))\nonumber\\
    =c^s&(\mathbf{g}_s\big([f_S(\delta_i,\zeta),\boldsymbol{\delta}_{-i}]\;;\;\mathbf{g}^*([\widetilde{\lambda}_i,\boldsymbol{\phi}_{-i}]),\boldsymbol{\pi}^*([\widetilde{\lambda}_i,\boldsymbol{\phi}_{-i}]),\omega\big)).
\end{align*}
Taking expectation on both sides of the equality with respect to $\omega\sim\mathbb{P}_{\boldsymbol{\pi}^*([\widetilde{\lambda}_i,\boldsymbol{\phi}_{-i}])}$ and using (\ref{cHatDefn}) gives
\begin{align*}
    \mathbb{E}_{\omega\sim\mathbb{P}_{\boldsymbol{\pi}^*([\widetilde{\lambda}_i,\boldsymbol{\phi}_{-i}])}}&c^s(\mathbf{g}_s(\boldsymbol{\delta};\mathbf{g}^*([\widetilde{\lambda}_i,\boldsymbol{\phi}_{-i}]),\boldsymbol{\nu},(\zeta,\omega)))\nonumber\\
    =\widehat{c^s}&([f_S(\delta_i,\zeta),\boldsymbol{\delta}_{-i}];\mathbf{g}^*([\widetilde{\lambda}_i,\boldsymbol{\phi}_{-i}]),\boldsymbol{\pi}^*([\widetilde{\lambda}_i,\boldsymbol{\phi}_{-i}]))).
\end{align*}
Consequently, 
\begin{align*}
    &\widehat{c^s}(\boldsymbol{\delta}\;;\;\mathbf{g}^*([\widetilde{\lambda}_i,\boldsymbol{\phi}_{-i}]),\boldsymbol{\nu})\nonumber\\
    &\coloneqq\mathbb{E}_{(\zeta,\omega)\sim\mathbb{P}_{\kappa}\times\mathbb{P}_{\boldsymbol{\pi}^*(\widetilde{\lambda}_i,\boldsymbol{\phi}_{-i})}}\big[c^s(\mathbf{g}_s(\boldsymbol{\delta};\mathbf{g}^*(\widetilde{\lambda}_i,\boldsymbol{\phi}_{-i}),\boldsymbol{\nu},(\zeta,\omega)))\big]\nonumber\\
    &\;\;\;=\mathbb{E}_{\zeta\sim\mathbb{P}_{\kappa}}\big[\widehat{c^s}([f_S(\delta_i,\zeta),\boldsymbol{\delta}_{-i}];\mathbf{g}^*([\widetilde{\lambda}_i,\boldsymbol{\phi}_{-i}]),\boldsymbol{\pi}^*([\widetilde{\lambda}_i,\boldsymbol{\phi}_{-i}])))\big],
\end{align*}
and so
\begin{align}
    \mathbb{E}_{\boldsymbol{\delta}\sim\mathbb{P}_{([\lambda_i,\boldsymbol{\phi}_{-i}])}}\widehat{c^s}(\boldsymbol{\delta}\;;\mathbf{g}^*(&[\widetilde{\lambda}_i,\boldsymbol{\phi}_{-i}]),\boldsymbol{\nu})\nonumber\\
    =\mathbb{E}_{\boldsymbol{\delta}'\sim\mathbb{P}_{([\widetilde{\lambda}_i,\boldsymbol{\phi}_{-i}])}}\big[\widehat{c^s}(&\boldsymbol{\delta}'\;;\mathbf{g}^*([\widetilde{\lambda}_i,\boldsymbol{\phi}_{-i}]),\boldsymbol{\pi}^*([\widetilde{\lambda}_i,\boldsymbol{\phi}_{-i}]))\big],\label{l2u4}
\end{align}
where we have once again used the fact that $f_S(\delta_i,\zeta)\sim\mathbb{P}_{\widetilde{\lambda}_i}$ and so $[f_S(\delta_i,\zeta),\boldsymbol{\delta}_{-i}]\sim\mathbb{P}_{([\widetilde{\lambda}_i,\boldsymbol{\phi}_{-i}])}.$

The desired result follows by noting that
\begin{align*}
    q^*(\lambda_i,\boldsymbol{\phi}_{-i})\leq\;\;\;\;\;\;\;\;\;\;\;\;\;\;\;\;\;\;\;\;\;\;\;\;\;\;\;\;\;\;\;\;\;\;\;\;\;\;\;\;\;\;\;\;&\nonumber\\
    c^g\big(\mathbf{g}^*([\widetilde{\lambda}_i,\boldsymbol{\phi}_{-i}])\big)+\mathbb{E}_{\boldsymbol{\delta}\sim\mathbb{P}_{(\lambda_i,\boldsymbol{\phi}_{-i})}}\bigg[\widehat{c^s}(\boldsymbol{\delta};\mathbf{g}^*&([\widetilde{\lambda}_i,\boldsymbol{\phi}_{-i}]),\boldsymbol{{\nu}})\nonumber\\
    -&\sum_{j=1}^{n_s}\widehat{\nu}_j(\delta_j,\boldsymbol{\delta})\bigg]\nonumber\\
    &\;=q^*(\widetilde{\lambda}_i,\boldsymbol{\phi}_{-i}),
\end{align*}
where the equality follows using (\ref{l2u2}) and (\ref{l2u4}).
\end{proof}

\section{Proof of Lemma 3}\label{Lemma3Proof}
\begin{proof}
Let $\psi$ be that element of $\Theta$ that has the distribution function $F_{\psi}(t)=\max\{F_{\theta_i}(t),F_{\phi_i}(t)\}.$ It is easy to see that
\begin{align}
    F_{\psi}(t)-F_{\phi_i}(t)\geq0\label{psiMinusPhigeqzero}
\end{align}
for all $t\in\{1,\hdots,T\},$ and that
\begin{align}
    \sup_{t\in\{1,\hdots,T\}}\{F_{\psi}(t)-F_{\phi_i}(t)\}=\alpha(\theta_i,\phi_i).\label{psiphismall}
\end{align}

Note that for all $t\in\{1,\hdots,T\},$ $\mathbb{P}_{\psi}(t)-\mathbb{P}_{\phi_i}(t)=F_{\psi}(t)-F_{\psi}(t-1)-\big(F_{\phi_i}(t)-F_{\phi_i}(t-1)\big)\leq\big\vert F_{\psi}(t)-F_{\phi_i}(t)\big\vert+\big\vert F_{\phi_i}(t-1)-F_{\psi}(t-1)\big\vert\leq2\alpha(\theta_i,\phi_i),$ where the last inequality follows from (\ref{psiMinusPhigeqzero}) and (\ref{psiphismall}). Therefore,
\begin{align}
    \sup_{t\in\{1,\hdots,T\}}\{\mathbb{P}_{\psi}(t)-\mathbb{P}_{\phi_i}(t)\}\leq 2\alpha(\theta_i,\phi_i).\label{Pdifferslittle}
\end{align}

{For $t=1,\hdots,T,$ define
\begin{align}
    \beta_{\boldsymbol{\phi},{i}}(t;\mathbf{g},\boldsymbol{\pi})\coloneqq \mathbb{E}_{\boldsymbol{\delta}\sim\mathbb{P}_{\boldsymbol{\phi}}}\big[\beta(\boldsymbol{\delta},\mathbf{g},\boldsymbol{\pi})\big\vert\delta_i=t\big]\label{betaDefn}
\end{align}
so that $\beta_{\boldsymbol{\phi},i}(t;\mathbf{g},\boldsymbol{\pi})$ denotes the conditional expected cost of satisfying the demand given that EV $i$ disconnects from the grid at time $t,$ the EV departure profiles is distributed according to $\mathbb{P}_{\boldsymbol{\phi}},$ the generator's energy dispatch sequence is $\mathbf{g},$ and the storage policy is $\boldsymbol{\pi}$. Note that the conditional expectation in (\ref{betaDefn}) is well defined for every $t$, thanks to (\ref{thetainTheta}). Note also that
\begin{align}
    q^*(\phi_i,\boldsymbol{\phi}_{-i})=\mathbb{E}_{\delta_i\sim\mathbb{P}_{\phi_i}}\big[\beta_{\boldsymbol{\phi},{i}}(\delta_i;\mathbf{g}^*(\boldsymbol{\phi}),\boldsymbol{\pi}^*(\boldsymbol{\phi}))\big].\label{qandbeta}
\end{align}}

Now,
\begin{align}
    \mathbb{E}_{\delta_i\sim\mathbb{P}_{\psi}}\big[\beta_{\boldsymbol{\phi},{i}}(\delta_i;\mathbf{g}^*(\boldsymbol{\phi}),\boldsymbol{\pi}^*(\boldsymbol{\phi}))\big]&\nonumber\\
    -\mathbb{E}_{\delta_i\sim\mathbb{P}_{\phi_i}}\big[\beta_{\boldsymbol{\phi},{i}}(\delta_i;\mathbf{g}^*(\boldsymbol{\phi}),\boldsymbol{\pi}^*(&\boldsymbol{\phi}))\big]\nonumber\\
    =\sum_{t=1}^{T}\big(\mathbb{P}_{\psi}(t)-\mathbb{P}_{\phi_i}(t)\big)\beta_{\boldsymbol{\phi},{i}}(&t;\mathbf{g}^*(\boldsymbol{\phi}),\boldsymbol{\pi}^*(\boldsymbol{\phi}))\nonumber\\
    \leq \Big[\sum_{t=1}^{T}(\mathbb{P}_{\psi}(t)-\mathbb{P}_{\phi_i}(t))^2\Big]^{\frac{1}{2}}\Big[\sum_{t=1}^{T}&\beta^2_{\boldsymbol{\phi},{i}}(t;\mathbf{g}^*(\boldsymbol{\phi}),\boldsymbol{\pi}^*(\boldsymbol{\phi}))\Big]^{\frac{1}{2}}\nonumber\\
    \leq 2\sqrt{T}\alpha(\theta_i,\phi_i)\Big[\sum_{t=1}^{T}&\beta^2_{\boldsymbol{\phi},{i}}(t;\mathbf{g}^*(\boldsymbol{\phi}),\boldsymbol{\pi}^*(\boldsymbol{\phi}))\Big]^{\frac{1}{2}},\label{smallgap1}
\end{align}
where the last inequality follows from (\ref{Pdifferslittle}). 
It follows from (\ref{qandbeta}) and the definitions of the functions $\mathbf{g}^*$ and $\boldsymbol{\pi}^*$ that
\begin{align}
    \mathbb{E}_{\delta_i\sim\mathbb{P}_{\psi}}\big[\beta_{\boldsymbol{\phi},{i}}&(\delta_i;\mathbf{g}^*([\psi,\boldsymbol{\phi}_{-i}]),\boldsymbol{\pi}^*([\psi,\boldsymbol{\phi}_{-i}]))\big]\nonumber\\
    &-\mathbb{E}_{\delta_i\sim\mathbb{P}_{\psi}}\big[\beta_{\boldsymbol{\phi},{i}}(\delta_i;\mathbf{g}^*(\boldsymbol{\phi}),\boldsymbol{\pi}^*(\boldsymbol{\phi}))\big]\leq 0.\label{smallgap2}
\end{align}
Adding (\ref{smallgap1}) and (\ref{smallgap2}) yields
\begin{align*}
    \mathbb{E}_{\delta_i\sim\mathbb{P}_{\psi}}\big[\beta_{\boldsymbol{\phi},{i}}(\delta_i;\mathbf{g}^*([\psi&,\boldsymbol{\phi}_{-i}]),\boldsymbol{\pi}^*([\psi,\boldsymbol{\phi}_{-i}]))\big]\nonumber\\
    -\mathbb{E}_{\delta_i\sim\mathbb{P}_{\phi_i}}&\big[\beta_{\boldsymbol{\phi},{i}}(\delta_i;\mathbf{g}^*(\boldsymbol{\phi}),\boldsymbol{\pi}^*(\boldsymbol{\phi}))\big]\nonumber\\
    \leq 2\sqrt{T}\alpha(\theta_i,\phi_i)&\Big[\sum_{t=1}^{T}\beta^2_{\boldsymbol{\phi},{i}}(t;\mathbf{g}^*(\boldsymbol{\phi}),\boldsymbol{\pi}^*(\boldsymbol{\phi}))\Big]^{\frac{1}{2}},
\end{align*}
which using (\ref{qandbeta}) implies
\begin{align}
    q^*(\psi,\boldsymbol{\phi}_{-i})-q^*(\phi_i,\boldsymbol{\phi}_{-i})\leq K\alpha(\theta_i,\phi_i)\label{lipschitzPropinterim}
\end{align}
where  
$$K\coloneqq\sup_{\boldsymbol{\phi}\in\Theta^{n_s}}2\sqrt{T}\Big[\sum_{t=1}^{T}\beta^2_{\boldsymbol{\phi},{i}}(t;\mathbf{g}^*(\boldsymbol{\phi}),\boldsymbol{\pi}^*(\boldsymbol{\phi}))\Big]^{\frac{1}{2}}.$$
That $K$ is finite follows from (\ref{costFinite}).

Since $F_{\psi}(t)\geq F_{\theta_i}(t)$ for all $t\in\{1,\hdots,T\},$ it follows from Lemma \ref{monotonic} that $q^*(\theta_i,\boldsymbol{\phi}_{-i})-q^*(\psi,\boldsymbol{\phi}_{-i})\leq0.$ Adding this inequality with (\ref{lipschitzPropinterim}) establishes (\ref{lipschitzProp}).

We now turn attention to the second part of the lemma. Let $\phi_i$ be such that (\ref{alphaDefn}) holds and denote by $t_0$ that element of $\{1,\hdots,T\}$ such that
\begin{align}
    F_{\theta_i}(t_0)-F_{\phi_i}(t_0)=\alpha(\theta_i,\phi_i).\label{Ft0isalpha}
\end{align}

We first have using the SLLN that $\mathbb{P}_{\theta_i}^\infty-$ almost surely,
\begin{align}
    \lim_{L\to\infty}\frac{1}{L}\sum_{l=1}^L\mathds{1}_{\{\delta_i(l)\leq t_0\}}=F_{\theta_i}(t_0).\label{deltaoccursoften1}
\end{align}

Now, let $\widehat{\delta}_i^\infty$ be any sequence such that $\sum_{l=1}^\infty\mathds{1}_{\{E_i(l,\widehat{\delta}_i^l,\phi_i\}}<\infty,$ i.e., $\{E_i(l,\widehat{\delta}_i^l,\phi_i)\}$ occurs only finitely often. Combining this with (\ref{Edefn}) implies the existence of $L_0$ such that  $\sum_{t=1}^{t_0}{\vert{f}_{i,t}(L,\widehat{\delta}_i^L,\phi_i)\vert < t_0r(L)}$ for all $L\geq L_0,$ which in turn implies that $\vert\sum_{t=1}^{t_0}{{f}_{i,t}(L,\widehat{\delta}_i^L,\phi_i)\vert < t_0r(L)}$ for all $L\geq L_0.$ Substituting (\ref{fdefn}) in this inequality and carrying out some algebra yields
\begin{align*}
    L[F_{\phi_i}(t_0)-t_0r(L)]<\sum_{l=1}^L&\mathds{1}_{\{\widehat{\delta}_i(l)\leq t_0\}}\nonumber\\ &<L[F_{\phi_i}(t_0)+t_0r(L)]
\end{align*}
for all $L\geq L_0.$ Using (\ref{rgoestozero}), this implies that
\begin{align}
    \lim_{L\to\infty}\frac{1}{L}\sum_{l=1}^L\mathds{1}_{\{\widehat{\delta}_i(l)\leq t_0\}}= F_{\phi_i}(t_0).\label{deltahatoccursoften1}
\end{align}
Since $\sum_{l=1}^L\mathds{1}_{\{\delta_i(l)<\widehat{\delta}_i(l)\}}\geq \sum_{l=1}^L\mathds{1}_{\{\delta_i(l)\leq t_0\}}-\sum_{l=1}^L\mathds{1}_{\{\widehat{\delta}_i(l)\leq t_0\}}$ for any $L\in\mathbb{Z}_+,$ dividing both sides of the inequality by $L$, taking the limit as $L\to\infty,$ and using (\ref{deltahatoccursoften1}) and (\ref{deltaoccursoften1}) yields

\begin{align*}
    \lim_{L\to\infty}\frac{1}{L}\sum_{l=1}^l\mathds{1}_{\{{\delta}_i(l)<\widehat{\delta}_i(l)\}}\geq F_{\theta_i}(t_0)-F_{\phi_i}(t_0)=\alpha(\theta_i,\phi_i)
\end{align*}
$\mathbb{P}_{\theta_i}^\infty-$ almost surely, where the equality follows from (\ref{Ft0isalpha}). This establishes (\ref{deadlinemissedoften}). 
\end{proof}

\section{Proof of Lemma 4}\label{Lemma4Proof}
\begin{proof}
Let $$\mathcal{I}_L\coloneqq\{l\leq L: \mathds{1}_{\{E_i(l,\widehat{\delta}_i^l,\phi_i)\}}=1\}.$$ Denote by $I_L^*$ the largest element of $\mathcal{I}_L,$ and note that if $$\sum_{l=1}^\infty\mathds{1}_{\{E_i(l,\widehat{\delta}_i^l,\phi_i)\}}=\infty,$$ then $I_L^*=L$ for infinitely many values of $L.$ Now, $$\frac{1}{L}\sum_{l=1}^{L} J_p(l)\mathds{1}_{\{E_i(l,\widehat{\delta}_i^l,\phi_i)\}}=\frac{1}{L}\sum_{l\in\mathcal{I}_{L}} J_p(l)\geq\frac{1}{L}J_p({I}^*_L),$$ and so $$\limsup_{L\to\infty}\frac{1}{L}\sum_{l=1}^{L} J_p(l)\mathds{1}_{\{E_i(l,\widehat{\delta}_i^l,\phi_i)\}}\geq\limsup_{L\to\infty}\frac{J_p(I_L^*)}{L}=\infty,$$ where the last equality follows from (\ref{JpOmegal}) and the fact that $I_L^*=L$ for infinitely many values of $L.$  
\end{proof}

\end{appendices}

\end{document}